\documentclass[preprint,preprintnumbers,amsmath,amssymb,floatfix,nofootinbib]{revtex4}
\usepackage{epsf}
\usepackage{graphicx}
\usepackage{subfigure}
\usepackage{color}
\begin{document}
\title{Exploring Two Higgs Doublet Models Through Higgs Production}
\author{Chien-Yi Chen and S.~Dawson}

\affiliation{
Department of Physics, Brookhaven National Laboratory, Upton, New York, 11973
\vspace*{.5in}}

\date{\today}

\begin{abstract}
We discuss the connections between the recently observed Higgs-like particle and rare  $B$ decays in the context of two Higgs
doublet models (2HDMs).  The measured decays of the Higgs boson to fermions and gauge bosons, along with the observation of the decay
$B_s\rightarrow \mu^+\mu^-$, place stringent restrictions on the allowed parameter space of 2 Higgs doublet models. Future measurements of $h^0\rightarrow
\gamma\gamma$ can potentially exclude type I 2HDMs,  while the parameters of other 2HDMs are already severely restricted.  The recent observations of
the $h^0\rightarrow \tau^+\tau^-$ and $h^0\rightarrow b {\overline b}$ decays  further constrain the models. 
\end{abstract}

\maketitle

\section{Introduction}
The discovery of a Higgs-like particle at the LHC with a mass  $m_{h^0}\sim 125~GeV$ is the beginning of the
exploration of the source of electroweak symmetry breaking.  The new particle has been observed in a number of different
final states with rates roughly consistent with the Standard Model predictions, although the decay rate to $\gamma\gamma$
is slightly high.  The crucial task now is to extract the
properties (spin/parity and couplings to fermions and gauge bosons) of this particle as precisely as possible and determine whether 
they correspond to those predicted by the minimal Standard Model.  It is also important to explore the possibility that
there is a spectrum of Higgs-like states.  

Many beyond the Standard Model  (BSM ) scenarios have a Higgs-like particle which has Standard Model -like couplings to fermions and gauge bosons in
the low energy limit, and precision measurements of the Higgs particle couplings can serve to limit the parameters of  BSM models.
A well motivated extension of the Standard Model is obtained by adding a second $SU(2)_L$ Higgs doublet,
leading to $5$  physical Higgs particles: $h^0$, $H^0$, $A^0$, and $H^\pm$.  We will consider the possibility that the particle observed
at the LHC is the lightest neutral Higgs particle,  $h^0$,  of a $2$ Higgs doublet model (2HDM.)\footnote{The possibility that the observed particle is the heavier neutral Higgs
particle, $H^0$,  of a 2HDM has been examined elsewhere\cite{Ferreira:2012my,Drozd:2012vf,Chang:2012ve}.}

The 2HDM models generically have
tree level flavor changing neutral
currents  (FCNCs)  from Higgs exchanges unless there is a global or discrete symmetry which forbids such
 interactions\cite{Branco:2011iw,Grossman:1994jb} and therefore  
we consider only
the class of models where there is a discrete $Z_2$ symmetry such that one of the fermions couples only to a single Higgs doublet.  FCNCs
are highly suppressed in this case and provide stringent limits on the parameters of the models.  There are four possibilities for 2HDMs of
this type which are
typically called the Type I, Type II, Lepton Specific, and Flipped models\cite{Branco:2011iw}.    We review the limits from FCNCs on these models, and determine the
range of parameters consistent with the latest Higgs measurements at the LHC.  
Our goal is to study the extent to which LHC measurements in the Higgs sector can restrict the possibilities for 2HDMs.
Previous works have examined
the possibility of enhancing the branching ratio $h^0\rightarrow \gamma\gamma$\cite{Ferreira:2011aa,Ferreira:2012my,Alves:2012ez}, 
and various other channels\cite{Craig:2012vn,Craig:2012pu,Bai:2012ex,Azatov:2012qz,Dobrescu:2012td} in the 2HDMs
and we discuss the implications of a measurement of $h^0\rightarrow \gamma\gamma$
which differs from the Standard Model prediction and demonstrate the impact of the  
$h^0\rightarrow b {\overline b}$ and $h^0\rightarrow \tau^+\tau^-$ measurements.   Finally, we show how the recent measurement of $B_s\rightarrow
\mu^+\mu^-$ \cite{:2012ct} serves to further restrict 2HDMs.

\section{Review of 2HDMs}
\begin{table}[t]
\caption{Neutral Higgs Couplings in the 2HDMs}
\centering
\begin{tabular}[t]{|c|c|c|c|c|}
\hline\hline
& I& II& Lepton Specific& Flipped\\
\hline
$g_{hVV}$ & $\sin(\beta-\alpha)$ & $\sin (\beta-\alpha)$ &$\sin (\beta-\alpha)$&$\sin (\beta-\alpha)$\\
$g_{ht\overline{t}}$&${\cos\alpha\over\sin\beta}$&${\cos\alpha\over\sin\beta}$&${\cos\alpha\over\sin\beta}$&${\cos\alpha\over\sin\beta}$\\
$g_{hb{\overline b}}$ &${\cos\alpha\over\sin\beta}$&$-{\sin\alpha\over\cos\beta}$&${\cos\alpha\over\sin\beta}$&$-{\sin\alpha\over \cos\beta}$\\
$g_{h\tau^+\tau^-}$&${\cos\alpha\over \sin\beta}$&$-{\sin\alpha\over \cos\beta}$&$-{\sin\alpha\over \cos\beta}$&${\cos\alpha\over\sin\beta}$\\
\hline
\end{tabular}
\label{table:coups}
\end{table}

\begin{table}[t]
\caption{Charged Higgs Couplings in the 2HDMs}
\centering
\begin{tabular}[t]{|c|c|c|c|c|}
\hline\hline
& I& II& Lepton Specific& Flipped\\
\hline
$\lambda_{tt}$ & $\cot\beta$ & $\cot\beta$ &$\cot\beta$&$\cot\beta$\\
$\lambda_{bb}$&$\cot\beta$ & $-\tan\beta$ &$\cot\beta$& $-\tan\beta$\\
$\lambda_{\tau\tau}$ &$\cot\beta$ & $-\tan\beta$ &$-\tan\beta$&$\cot\beta$\\
\hline
\end{tabular}
\label{tab_ch}
\end{table}

The Higgs sector of the 2HDMs is parameterized in terms of $\tan\beta\equiv{v_2\over v_1}$, where $v_1$ and $v_2$
are the vacuum expectation values of the two Higgs doublets and satisfy $M_W^2=g^2(v_1^2+v_2^2)/2$, and the angle $\alpha$ which
diagonalizes the neutral Higgs mass matrix.  
The couplings of the lightest Higgs boson, $h^0$, to the Standard Model particles are parameterized as:
\begin{equation}
{\cal L}=-\Sigma_i g_{iih} {m_i\over v} {\overline f}_i f_ ih^0 -\Sigma_{i=W,Z}g_{hVV}{2M_V^2\over v} V_\mu V^\mu h^0\, ,
\end{equation}
where $g_{iih}=g_{hVV}=1$ in the Standard Model and $v=246GeV$. 
The $h^0$ coupling to gauge bosons is the same for all four models considered here, while the couplings to fermions 
differentiates between the models.
The Higgs Yukawa couplings normalized to their Standard Model values are summarized in Table \ref{table:coups}.  The Standard Model couplings
are obtained for $\sin(\beta-\alpha)=1$, $\sin \alpha=-\cos\beta$, and $\cos\alpha=\sin\beta$.
The charged Higgs-fermion couplings can be written as, 
\begin{equation}
{\cal L}={g\over \sqrt{2}M_W}
{\overline{t}}
\biggl(\lambda_{tt}m_t P_L-\lambda_{bb}m_b P_R\biggr)b H^+ -
{g\over \sqrt{2}M_W}{\overline{\nu}}\lambda_{ll} m_l P_Rl H^+ + {\rm {h.c.}}
\end{equation}
where $P_{L,R}={1\mp\gamma_5\over 2}$, $m_l$ is the charged lepton mass
 and the coefficients $\lambda_{ff}$ are given in Table \ref{tab_ch}.

Perturbative unitarity of the Yukawa couplings  is violated if
\begin{equation}
(y_i)^2=\biggl({g_{iih}\over\sqrt{2}}\biggr)^2~>~4\pi\, .
\end{equation}
For all models considered here this requires $\tan\beta > 0.28$.  Model II and the Flipped model require $\tan\beta < 140$, while the Lepton-specific
model has the limit $\tan\beta < 350$ from perturbative unitarity.   Other bounds on perturbative unitarity have been derived by requiring that the quartic
couplings in the scalar potential remain positive up to a high scale\cite{Kanemura:1999xf} 
and  by considering the perturbative unitarity of gauge boson scattering\cite{Kanemura:1993hm}.  These
bound typically give much lower upper bounds on $\tan \beta$ than those derived from perturbative unitarity of the Yukawa couplings, but we will not
consider them further since these limits can be evaded by postulating new physics at some high scale. 

There are also strong bounds on the 2HDMs from precision electroweak measurements, which can be parameterized by
the oblique parameters, $S,T$ and $U$.
In the limit, $M_{H^0}, M_A, M_{H^\pm}>>M_Z$ and subtracting the Standard Model contribution\cite{Haber:1999zh,He:2001tp},
\begin{eqnarray}
\alpha\Delta T &=& {1\over 16 \pi^2 v^2} \biggl\{f(M_A^2,M_{H^\pm}^2)-\sin^2(\beta-\alpha)\biggl(f(M_{H^0}^2,M_A^2)
-f(M_{H^0}^2,M_{H^\pm}^2)\biggr)
\nonumber \\
&&-\cos^2(\beta-\alpha)\biggl(
(f(M_{h^0}^2,M_A^2)-f(M_{h^0}^2,M_{H^\pm}^2)\biggr)\biggr\}
+{\cal{O}}\biggl(
{M_Z^2\over M_{H^0}^2},
{M_Z^2\over M_A^2}, 
{M_Z^2\over M_{H^\pm}^2}\biggr)\, ,
\nonumber \\
\Delta S&=& {1\over 12\pi}
\biggl\{ 
\cos^2(\beta-\alpha)
\biggl[
\log\biggl({M_{H^0}^2M_A\over M_{H^\pm}^2M_{h^0}}\biggr)
+g(M_{h^0}^2,M_A^2)\biggr]
\nonumber \\
&& +\sin^2(\beta-\alpha)
\biggl[g(M_{H^0}^2,M_A^2)-
\log\biggl({M_{H^\pm}^2\over M_{H^0}M_A}\biggr)\biggr]\biggr\}
+{\cal{O}}\biggl(
{M_Z^2\over M_{H^0}^2},
{M_Z^2\over M_A^2}, 
{M_Z^2\over M_{H^\pm}^2}\biggr)\, ,
\end{eqnarray} 
where
\begin{eqnarray}
f(x,y) & =& {x^2+y^2\over 2}-{x^2y^2\over x^2-y^2}\log{x^2\over y^2}\nonumber \\
g(x,y)&=& -{5\over 6}
+{2xy\over (x-y)^2}
+{(x+y)(x^2-4xy+y^2)\over 2 (x-y)^3}\log\biggl({x\over y}\biggr)\, .
\end{eqnarray}
For $M_A=M_{H^0}=M_{H^\pm}$ and $M_A>>M_Z$,
\begin{eqnarray}
\alpha\Delta T&\sim&{\cal O}\biggl({M_Z^2\over M_A^2}\biggr)
\nonumber \\
\Delta S&\sim& {1\over 12 \pi}\cos^2(\beta-\alpha) 
\biggl[\log\biggl({M_A^2\over M_{h^0}^2}\biggr)-{5\over 6}\biggr]\, ,
\end{eqnarray}
For $M_A=M_{H^0}=M_{H^\pm}\sim 1~TeV$, $\Delta T$ provides no useful limit on $\sin(\beta-\alpha)$,
while $\Delta S$ requires $\cos(\beta-\alpha) \sim 1$.
The  most significant restrictions come from $B$ decays, as discussed in the next section.

The branching ratios to fermions in the 2HDMs 
are simply scaled from the couplings of Table \ref{table:coups} and the total $h^0$ decay
width.  We use the Standard
Model branching ratios from the LHC Higgs cross section working group
for $M_{h^0}=125~GeV$\cite{Dittmaier:2011ti}: 
\begin{eqnarray}
\Gamma_{h^0}^{SM}&=&4.07 ~MeV  \nonumber \\
BR(h^0\rightarrow b{\overline b})^{SM}&=&.577\nonumber \\
BR(h^0\rightarrow \tau^+\tau^-)^{SM}&=&.063\nonumber \\
BR(h^0\rightarrow c{\overline c})^{SM}&=&.029\nonumber \\
BR(h^0\rightarrow W^+ W^-)^{SM}&=&.215\nonumber \\
BR(h^0\rightarrow ZZ)^{SM}&=&.026\nonumber \\
BR(h^0\rightarrow \gamma\gamma)^{SM}&=&2.28\times 10^{-3}\nonumber \\
BR(h^0\rightarrow gg)^{SM}&=&.086\,  .
\end{eqnarray}
The total widths in the 2HDMs are  given by,\footnote{We assume $g_{hWW}=g_{hZZ}$.}
\begin{eqnarray}
{\Gamma_{h^0}^{2HDM}\over\Gamma_{h^0}^{SM}}&\sim &
.577 g_{hbb}^2 + .029 g_{hcc}^2+
.063g_{h\tau\tau}^2+.241\sin^2(\beta-\alpha)
\nonumber \\ &&
+{\Gamma(h^0\rightarrow\gamma\gamma)^{2HDM}\over
\Gamma(h^0\rightarrow\gamma\gamma)^{SM}}
+
{\Gamma(h^0\rightarrow gg)^{2HDM}\over
\Gamma(h^0\rightarrow gg)^{SM}}+...
\end{eqnarray}
where we neglect other contributions  which are smaller than the $h^0\rightarrow \gamma\gamma$ branching ratio
and assume that there is no new physics beyond the 2HDM. 

The decay width to $\gamma\gamma$ is found using the exact form factors of
 Refs. \cite{Gunion:1989we,Djouadi:2005gj}.
For $M_{h^0}=125~GeV$, the dominant contributions are 
\begin{eqnarray}
{[\Gamma(h^0\rightarrow \gamma\gamma)]^{2HDM}\over
[\Gamma(h^0\rightarrow\gamma\gamma)]^{SM}} &\sim&
\biggl\{.28g_{htt}-.004g_{hbb}-.0036g_{h\tau\tau}
-1.27g_{hWW}\biggr\}^2
\nonumber \\ &&
+\biggl\{.0057g_{hbb}+.0033g_{h\tau\tau}\biggr\}^2\, .
\end{eqnarray}
 The tri-linear $h^0H^+H^-$ coupling
 is not enhanced by
a mass factor, and we neglect its negligible contribution\cite{Gunion:2002zf}.

Similarly, the decay to gluons proceeds predominantly through top and bottom quark loops,
\begin{equation}
{\Gamma(gg\rightarrow h^0)^{2HDM}\over
\Gamma(gg\rightarrow h^0)^{SM}}
=\biggl\{1.06g_{htt}-.06g_{hbb}\biggr\}^2+\biggl\{.086 g_{hbb}\biggr\}^2
\, .
\end{equation}
For large $\tan\beta$, the $b$ quark loop can contribute significantly to the $gg\rightarrow h^0$
production channel in Model II and the Flipped model.  In our numerical studies, we include
both the $b$ and $t$ contributions exactly in all cases .  

Using the results given above, the production from gluon fusion and the  following decay to $\gamma\gamma$ or to
$f{\overline f}$ pairs
can be described by,
\begin{eqnarray}
R_{\gamma\gamma}^{ggF}&=&{[\sigma(gg\rightarrow h^0)BR(h^0\rightarrow \gamma\gamma)]^{2HDM}
\over
[\sigma(gg\rightarrow h^0)BR(h^0\rightarrow \gamma\gamma)]^{SM}}\nonumber \\
R_{ff}^{ggF}&=&{[\sigma(gg\rightarrow h^0)BR(h^0\rightarrow f{\overline f})]^{2HDM}
\over
[\sigma(gg\rightarrow h^0)BR(h^0\rightarrow f{\overline f})]^{SM}}
\, .
\end{eqnarray}

\begin{figure}[tb]
\subfigure[]{
      \includegraphics[width=0.36\textwidth,angle=0,clip]{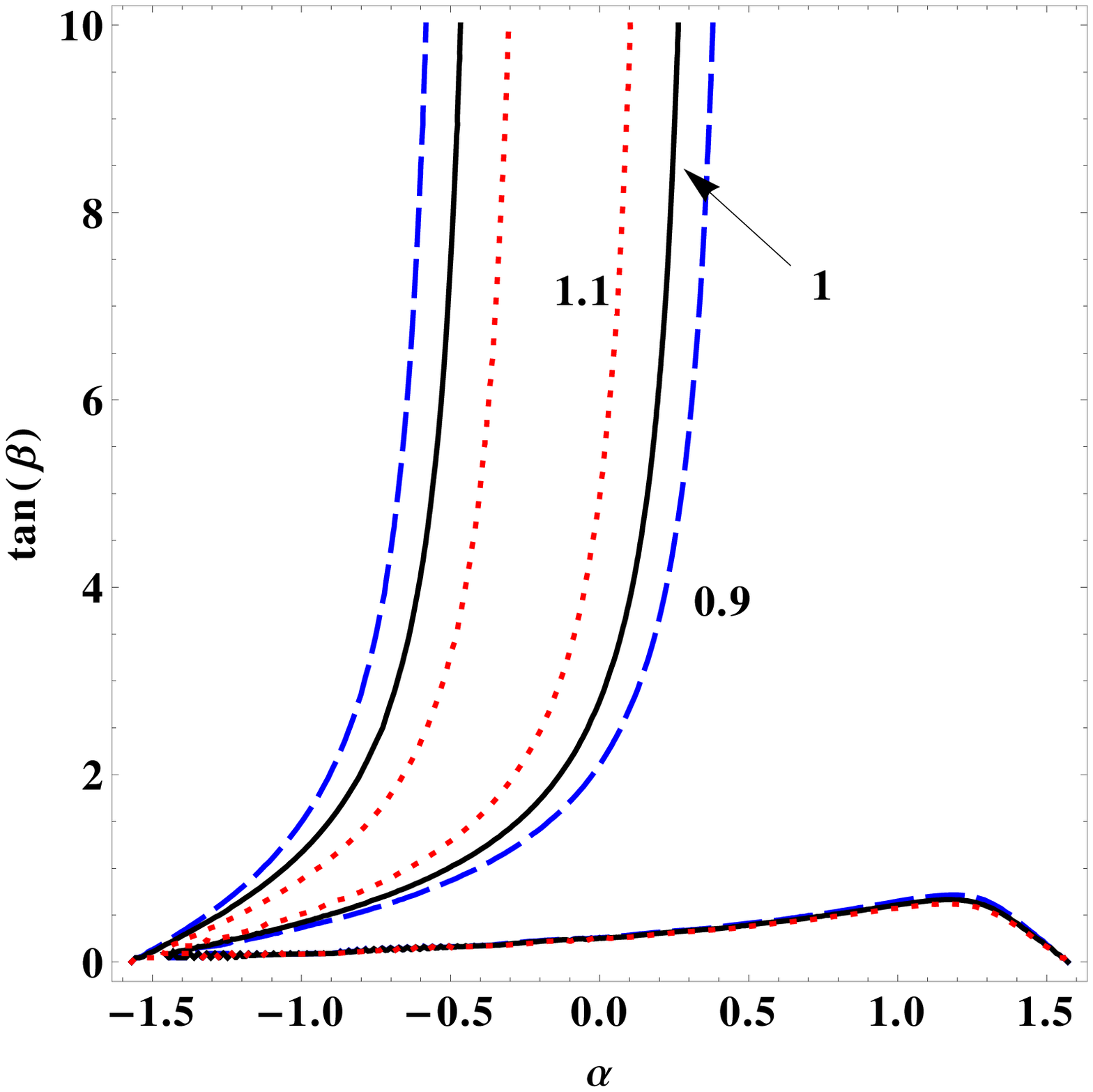}
}
\subfigure[]{
      \includegraphics[width=0.36\textwidth,angle=0,clip]{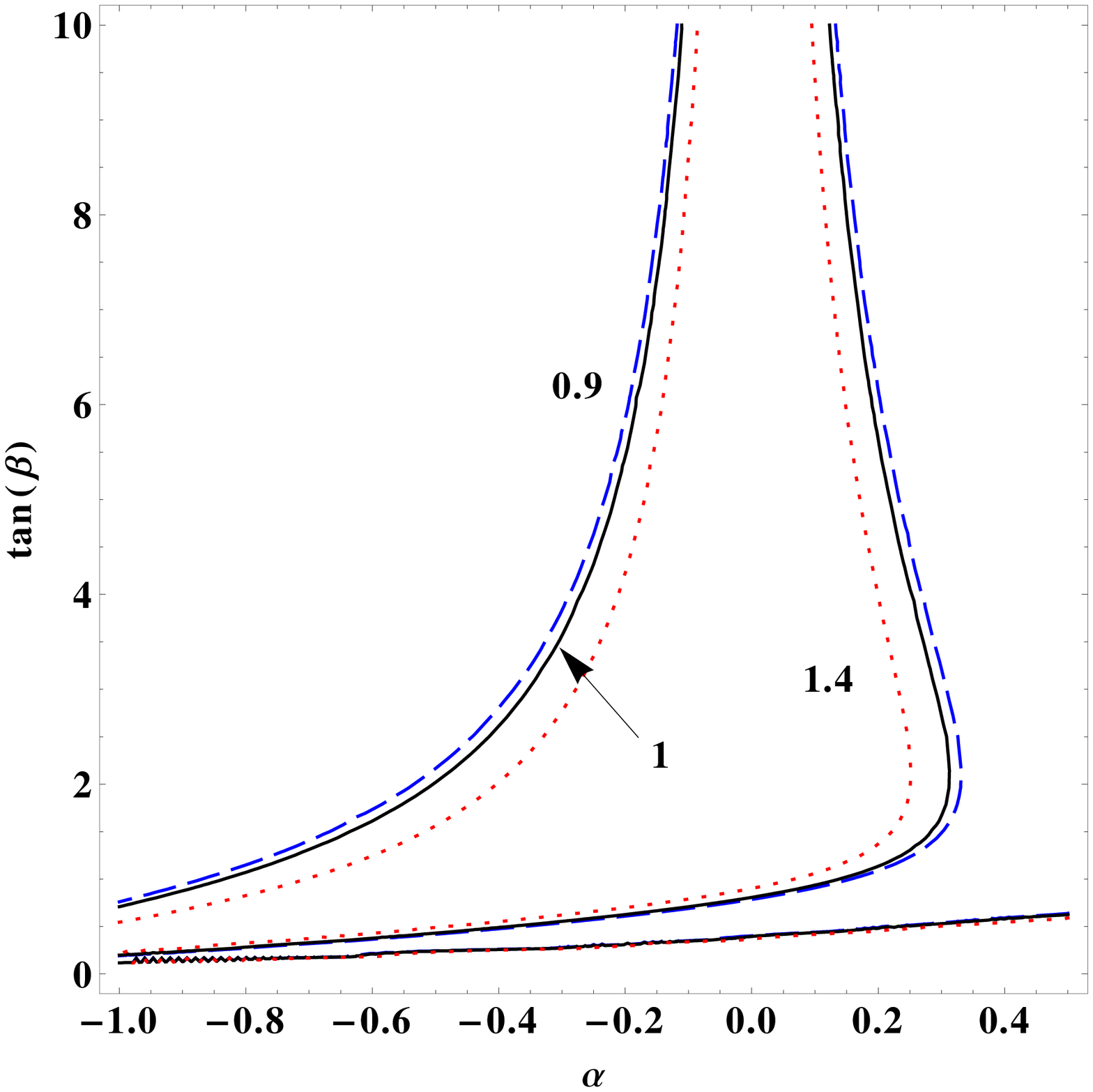}
}
\subfigure[]{
      \includegraphics[width=0.36\textwidth,angle=0,clip]{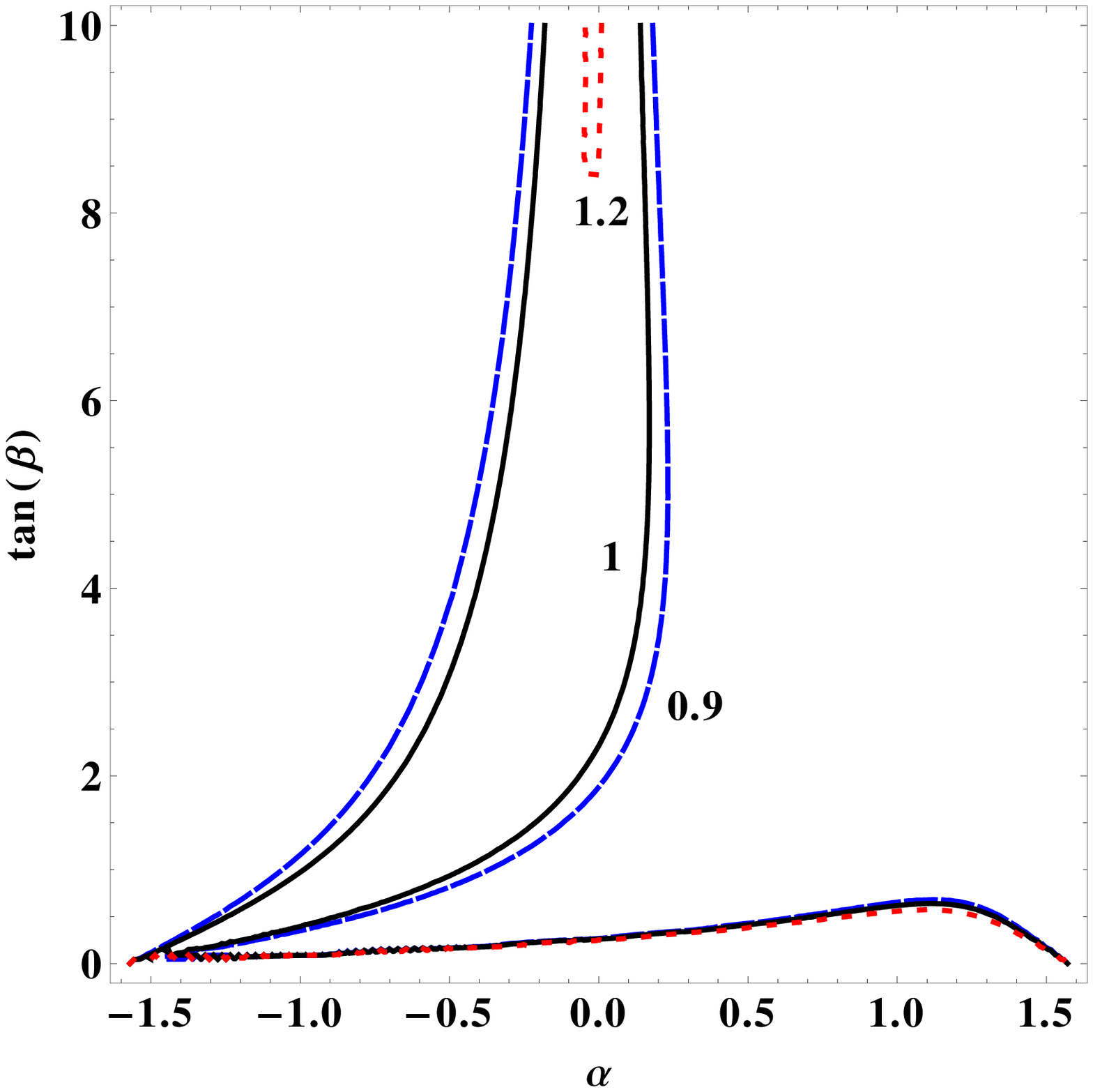}
}
\subfigure[]{
      \includegraphics[width=0.36\textwidth,angle=0,clip]{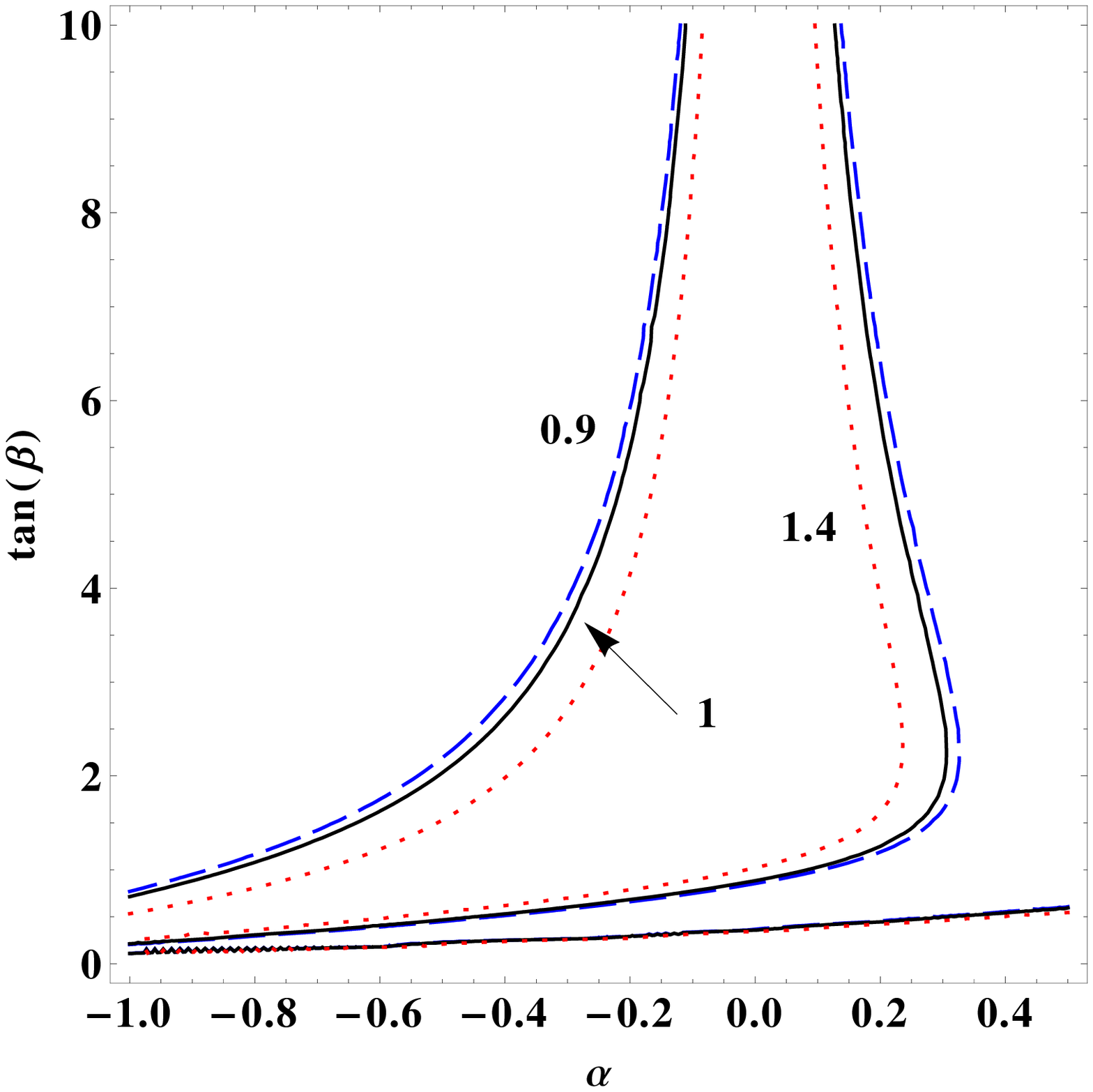}
}
\caption{ The contours correspond to constant $R^{ggF}_{\gamma\gamma}$ as defined in the text
for $\sqrt{S}=8~TeV$.  For Type I (a): 
The blue (dashed), black (solid), and red (dotted) lines are $R^{ggF}_{\gamma\gamma}=.9,1,1.1$, respectively.
 For Type II (b): 
The blue, black, and red lines are $R^{ggF}_{\gamma\gamma}=.9,1., 1.4$, respectively.
For the Lepton Specific model (c): 
The blue, black, and red lines are $R^{ggF}_{\gamma\gamma}=.9,1., 1.2$, respectively.
 For the Flipped model(d): 
The blue, black, and red lines are $R^{ggF}_{\gamma\gamma}=.9,1., 1.4$, respectively.
}
\label{hgg_fig}
\end{figure}
\begin{figure}[tb]
\subfigure[]{
      \includegraphics[width=0.36\textwidth,angle=0,clip]{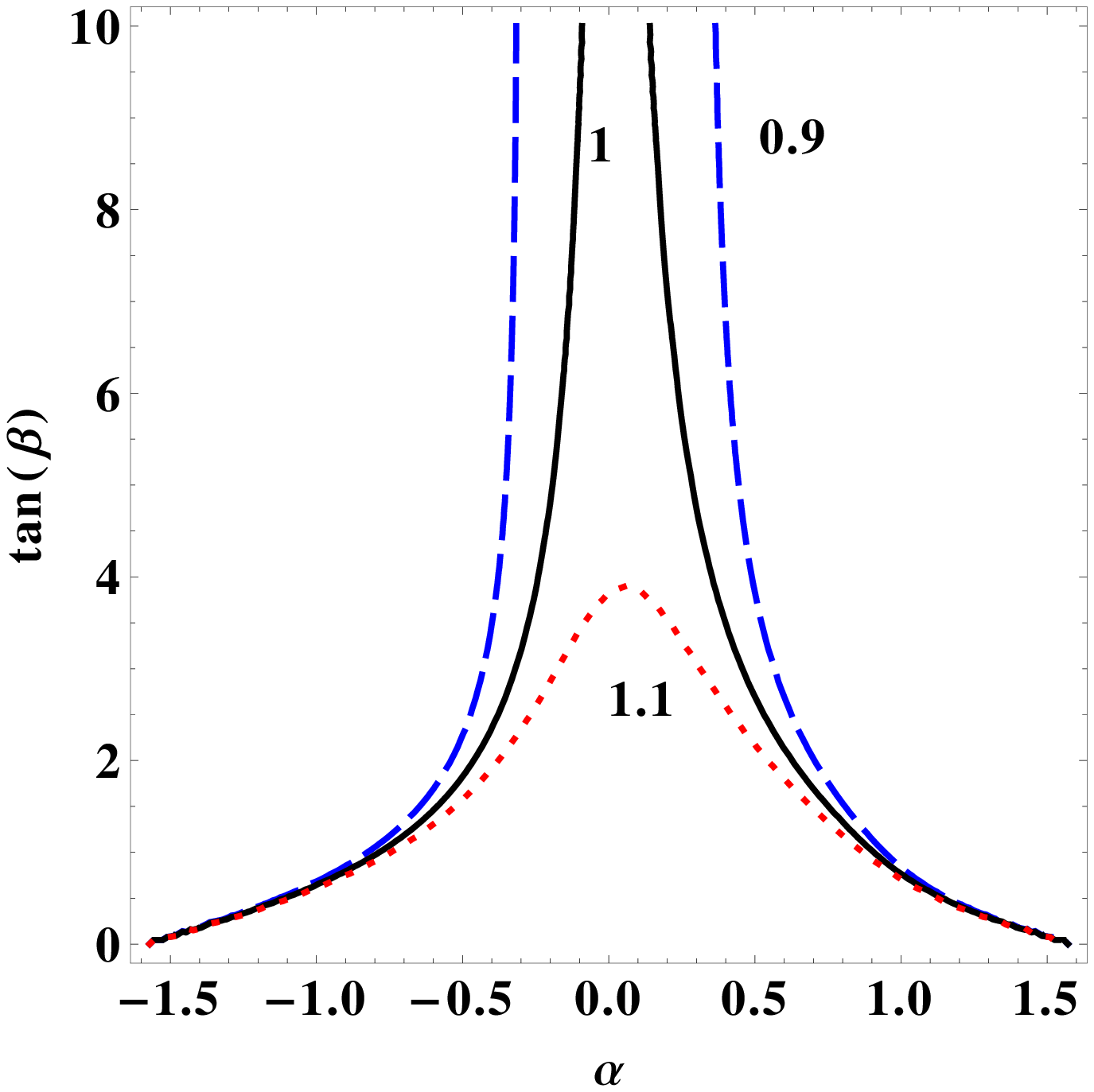}
}
\subfigure[]{
      \includegraphics[width=0.36\textwidth,angle=0,clip]{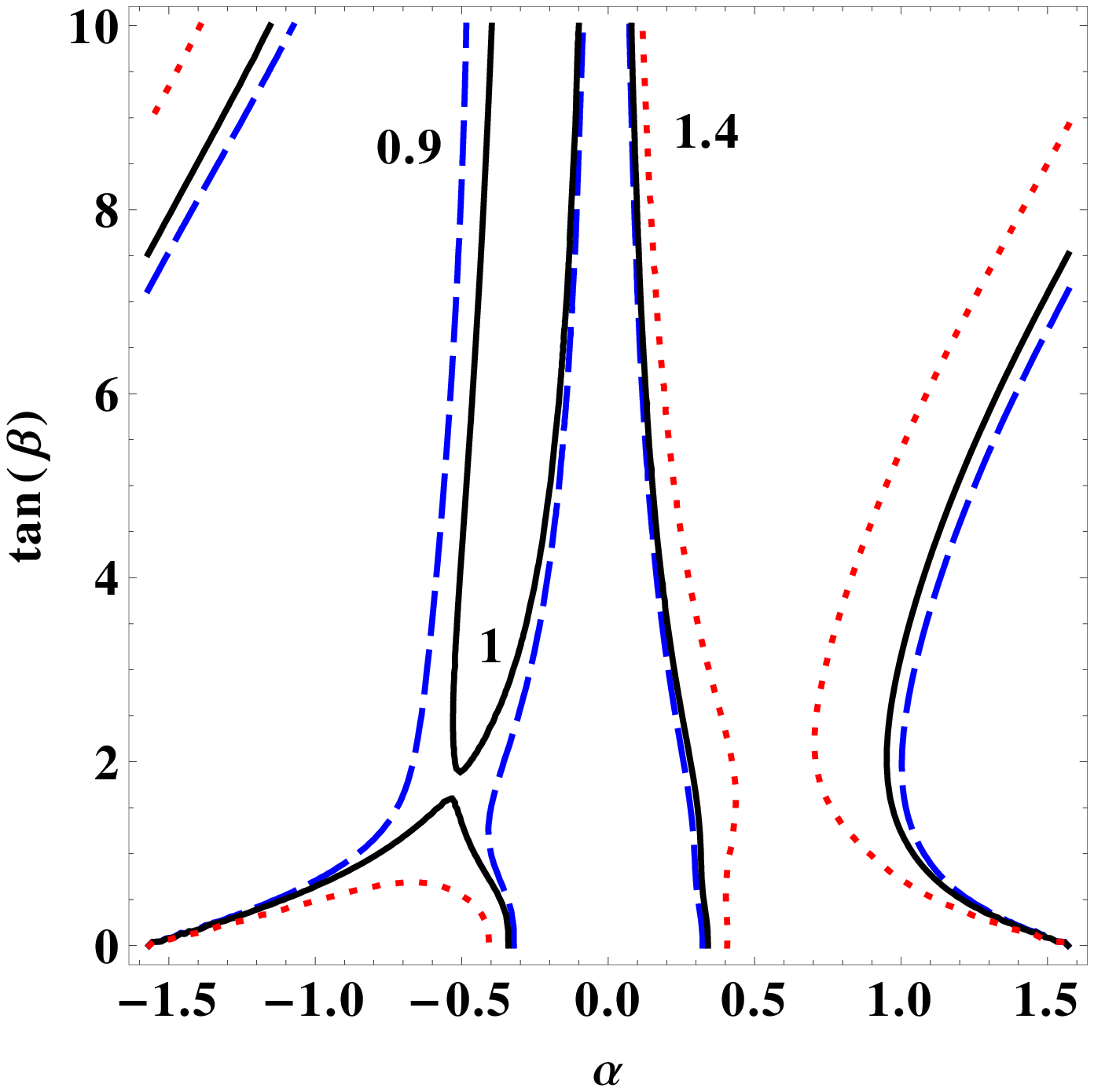}
}
\subfigure[]{
      \includegraphics[width=0.36\textwidth,angle=0,clip]{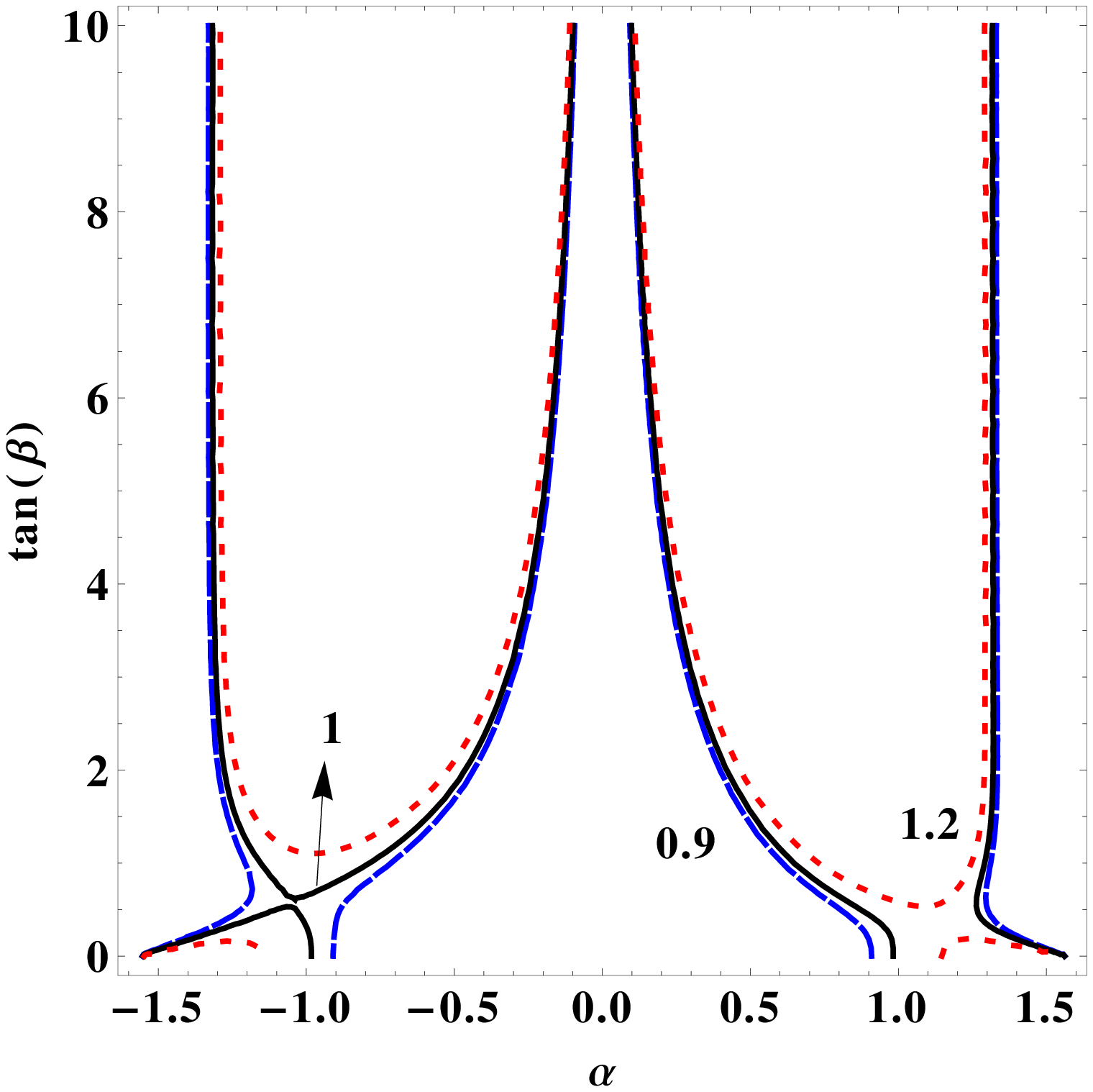}
}
\subfigure[]{
      \includegraphics[width=0.36\textwidth,angle=0,clip]{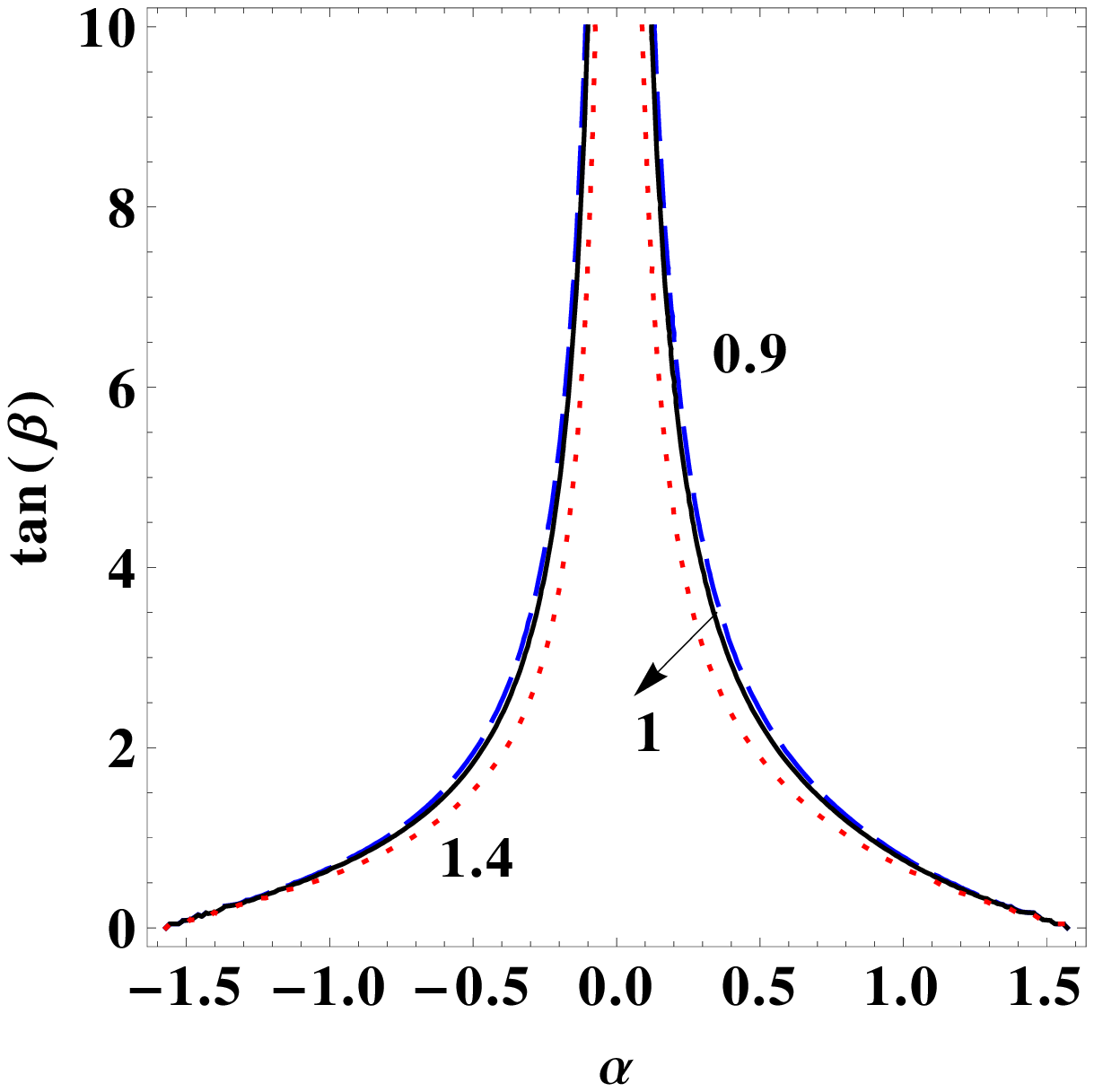}
}
\caption{
The contours correspond to constant  $R_{\tau\tau}^{ggF}$ as defined in the text
for $\sqrt{S}=8~TeV$.  For Type I (a): 
The blue (dashed), black (solid), and red (dotted) lines are $R_{\tau\tau}^{ggF}=.9,1,1.1$, respectively.
 For Type II (b): 
The blue, black, and red lines are $R_{\tau\tau}^{ggF}=.9,1., 1.4$, respectively.
For the Lepton-Specific model (c): 
The blue, black, and red lines are $R_{\tau\tau}^{ggF}=.9,1., 1.2$, respectively.
 For the Flipped Model(d): 
The blue, black, and red lines are $R_{\tau\tau}^{ggF}=.9,1., 1.4$, respectively.
}
\label{htt_ggf_fig}
\end{figure}

\begin{figure}[tb]
\subfigure[]{
      \includegraphics[width=0.36\textwidth,angle=0,clip]{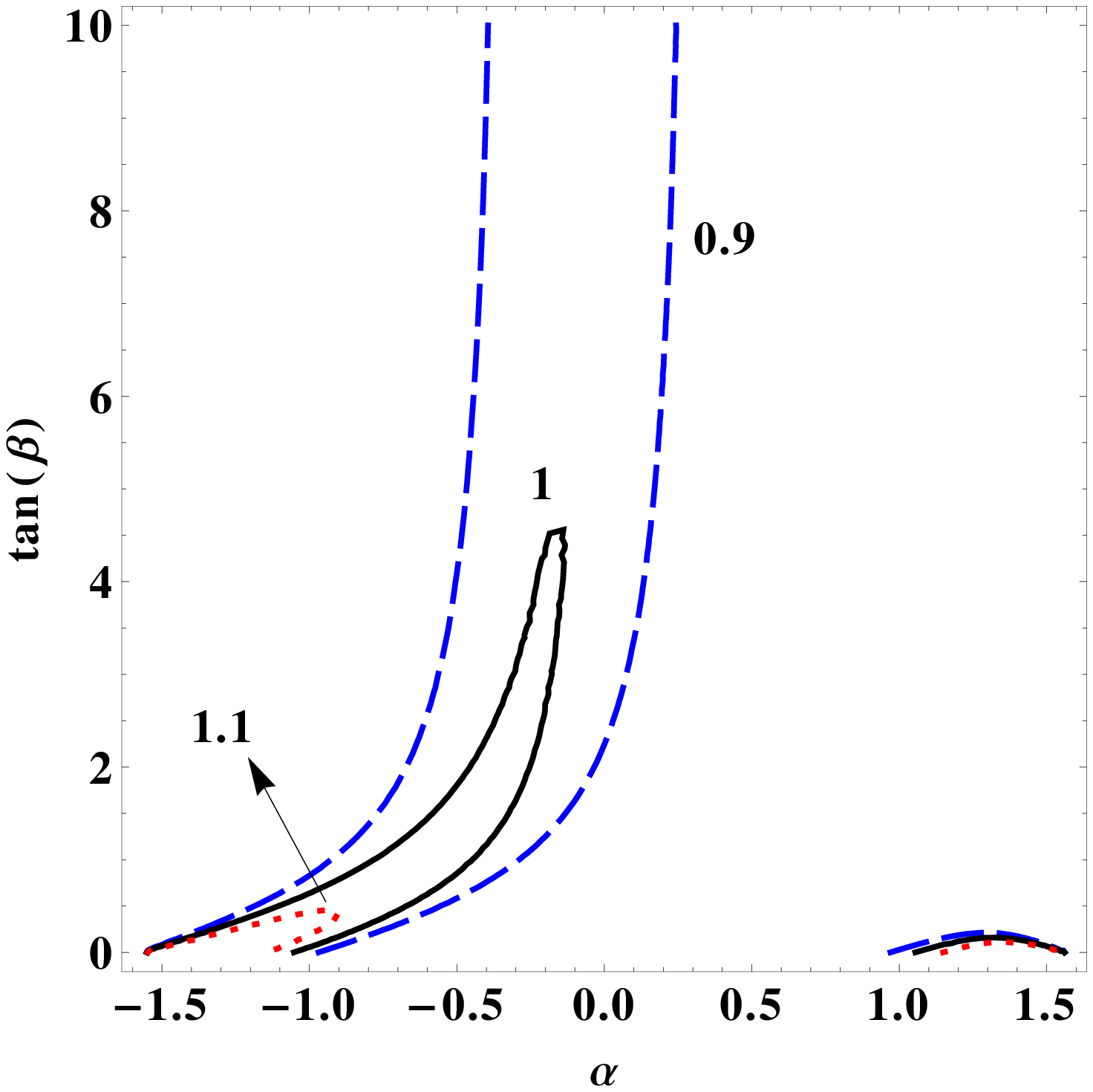}
}
\subfigure[]{
      \includegraphics[width=0.36\textwidth,angle=0,clip]{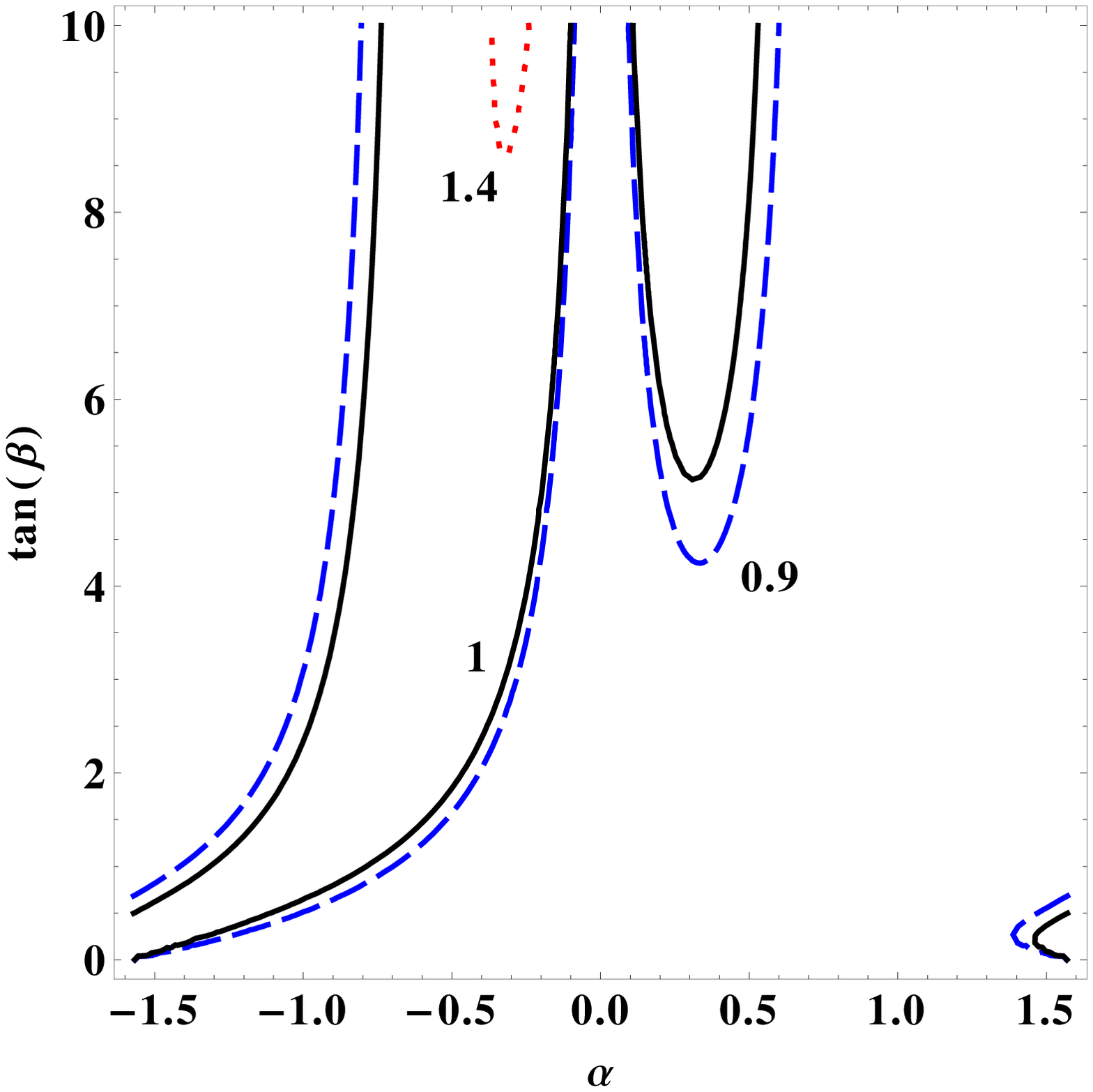}
}
\subfigure[]{
      \includegraphics[width=0.36\textwidth,angle=0,clip]{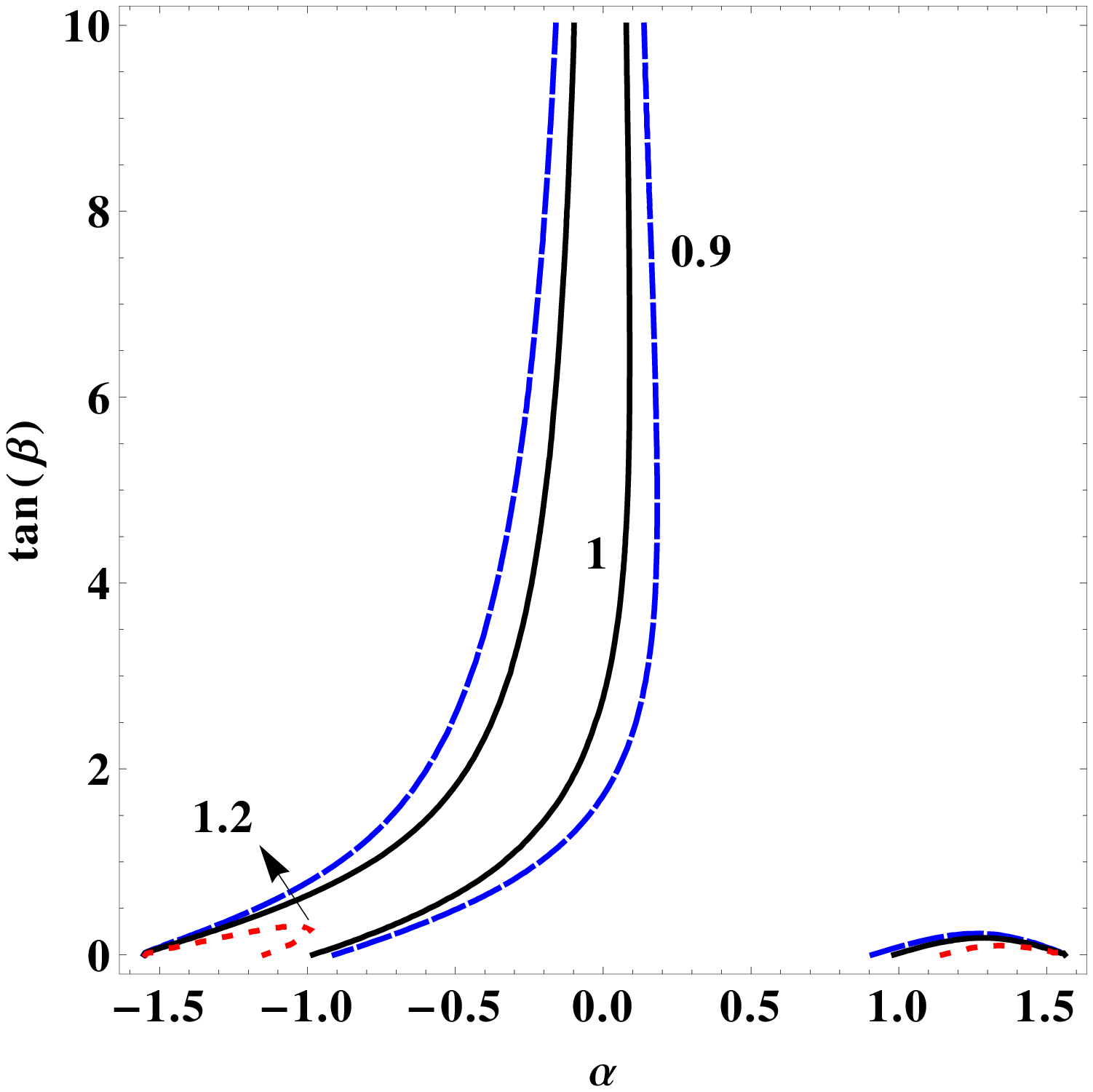}
}
\subfigure[]{
      \includegraphics[width=0.36\textwidth,angle=0,clip]{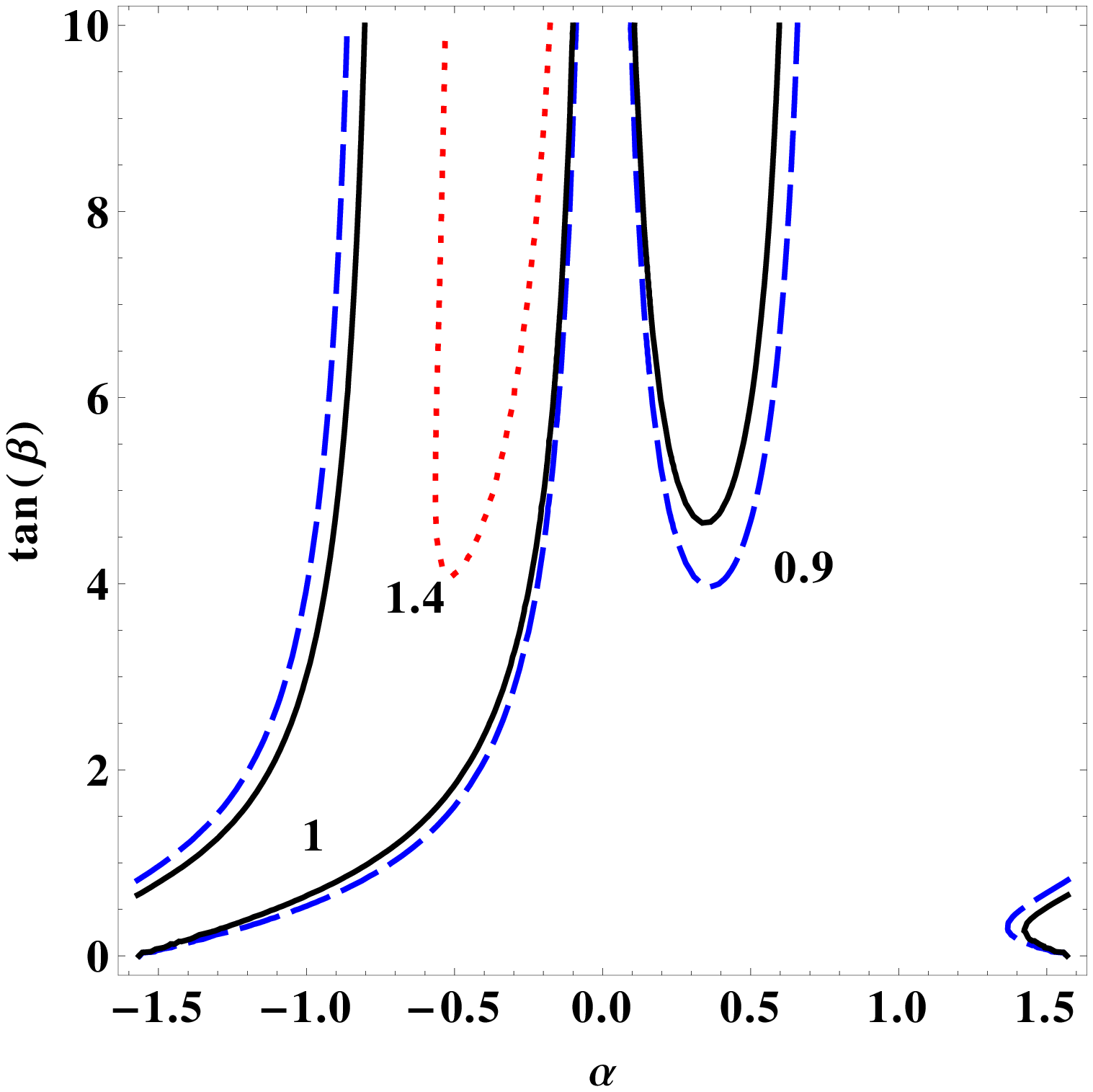}
}
\caption{
The contours correspond to constant  $R_{bb}^{VBF,Vh}$ as defined in the text.  For Type I (a): 
The blue (dashed), black (solid), and red (dotted) lines are $R_{bb}^{VBF,Vh}=.9,1,1.1$, respectively.
 For Type II (b): 
The blue, black, and red lines are $R_{bb}^{VBF,Vh}=.9,1., 1.4$, respectively.
For the Lepton-Specific model (c): 
The blue, black, and red lines are $R_{bb}^{VBF,Vh}=.9,1., 1.2$, respectively.
 For the Flipped Model(d): 
The blue, black, and red lines are $R_{bb}^{VBF,Vh}=.9,1., 1.4$, respectively.
}
\label{hbb_fig}
\end{figure}
The Higgs signal from gluon fusion with the subsequent decay to $\gamma\gamma$  is
shown in Fig. \ref{hgg_fig} for $\sqrt{S}=8~TeV$\footnote{There is very little change going from
$\sqrt{S}=7~TeV$ to $\sqrt{S}=8~TeV$.}. It is interesting to note that in Model I, it is not possible to obtain
$R_{\gamma\gamma}^{ggF}$ larger than $\sim 1.2$, so a future measurement which confirms the current
deviation of $R_{\gamma\gamma}^{ggF}$ from $1$ could serve to exclude Model I\footnote{This has also been
noted in Ref. \cite{Drozd:2012vf}.}.  For $R_{\gamma\gamma}^{ggF}>1$, only a narrow region of $\alpha\sim -0.5$
can be obtained.  The Lepton Specific model is similar to Model I at small $\tan\beta$, while at
large $\tan\beta$ ( for fixed non-zero $\alpha$) the total width is enhanced by the large branching ratio to $\tau^+\tau^-$, which decreases
$R_{\gamma\gamma}^{ggF}$.  A value of $R_{\gamma\gamma}^{ggF}\sim 1.2$ is only consistent
with the Lepton Specific model for $\alpha\sim 0$ and $\tan\beta > 8$.  Model II and the Flipped model have similar
predictions for $R_{\gamma\gamma}^{ggF}$ with a wide range of $\alpha$ and  $\tan\beta$ values 
consistent with $R_{\gamma\gamma}^{ggF}\sim 1$.

The gluon fusion production with decay to $\tau^+ \tau^-$ is shown in Fig. \ref{htt_ggf_fig}.  In Model I and the Flipped model, the
Standard Model rate can be obtained for small $\alpha$, relatively independently of $\tan\beta$.  Model II and the Lepton Specific
model only approximate the Standard Model rate for very specific values of $\alpha$ and $\tan\beta$ and hence a precise
measurements of $R_{\tau\tau}^{ggF}$ can serve to restrict these models significantly. 

\begin{figure}[tb]
\subfigure[]{
      \includegraphics[width=0.36\textwidth,angle=0,clip]{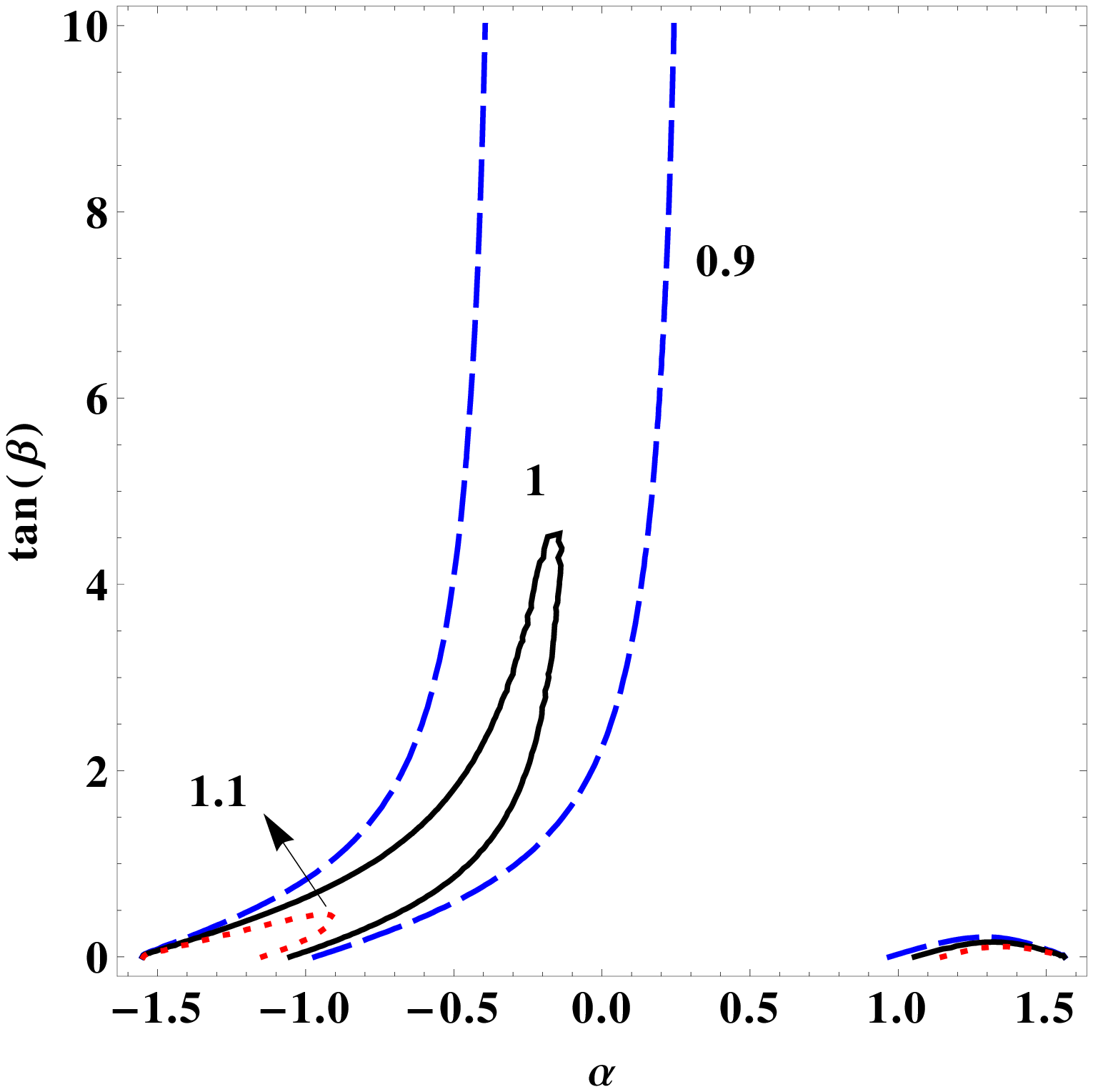}
}
\subfigure[]{
      \includegraphics[width=0.36\textwidth,angle=0,clip]{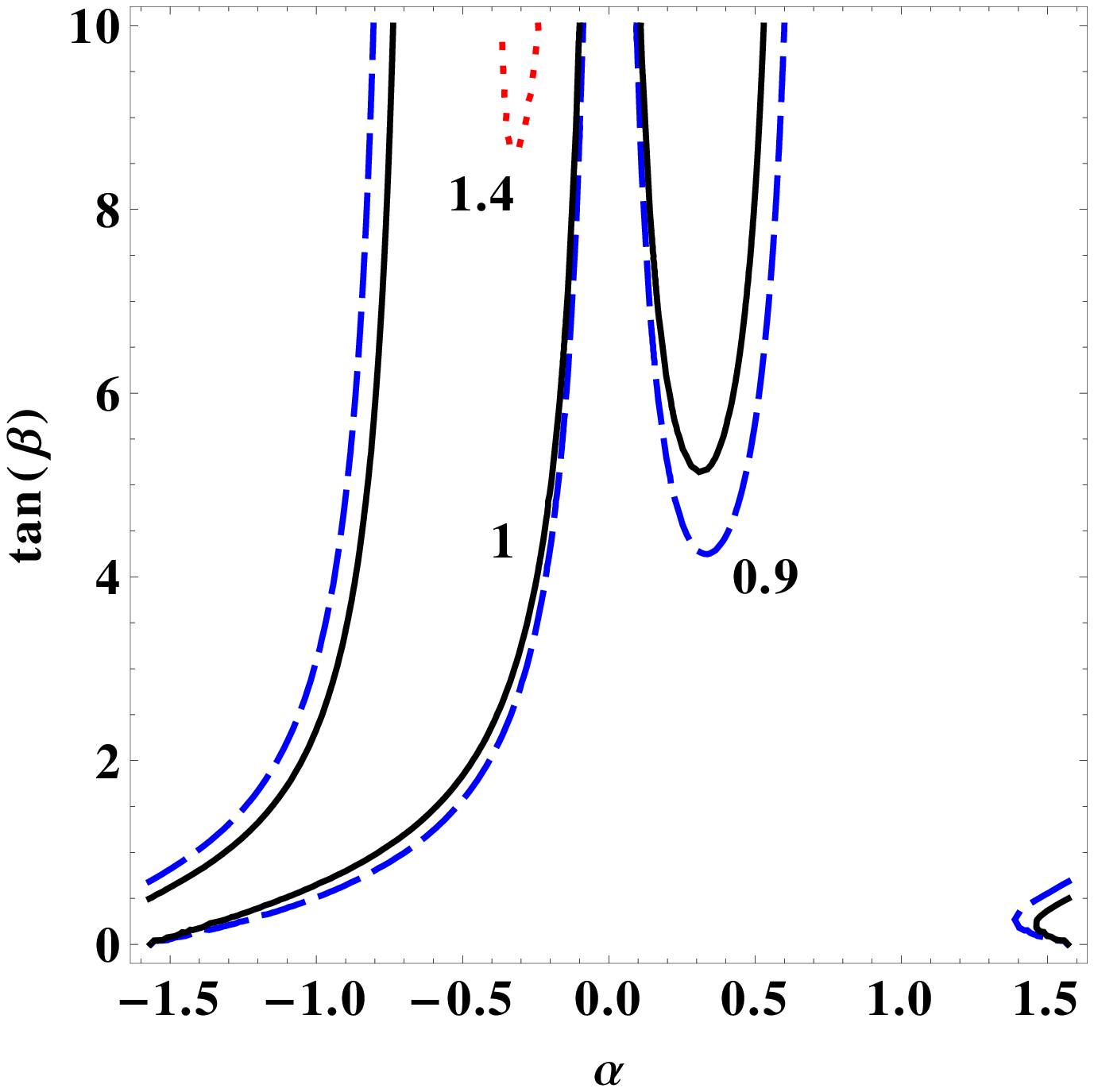}
}
\subfigure[]{
      \includegraphics[width=0.36\textwidth,angle=0,clip]{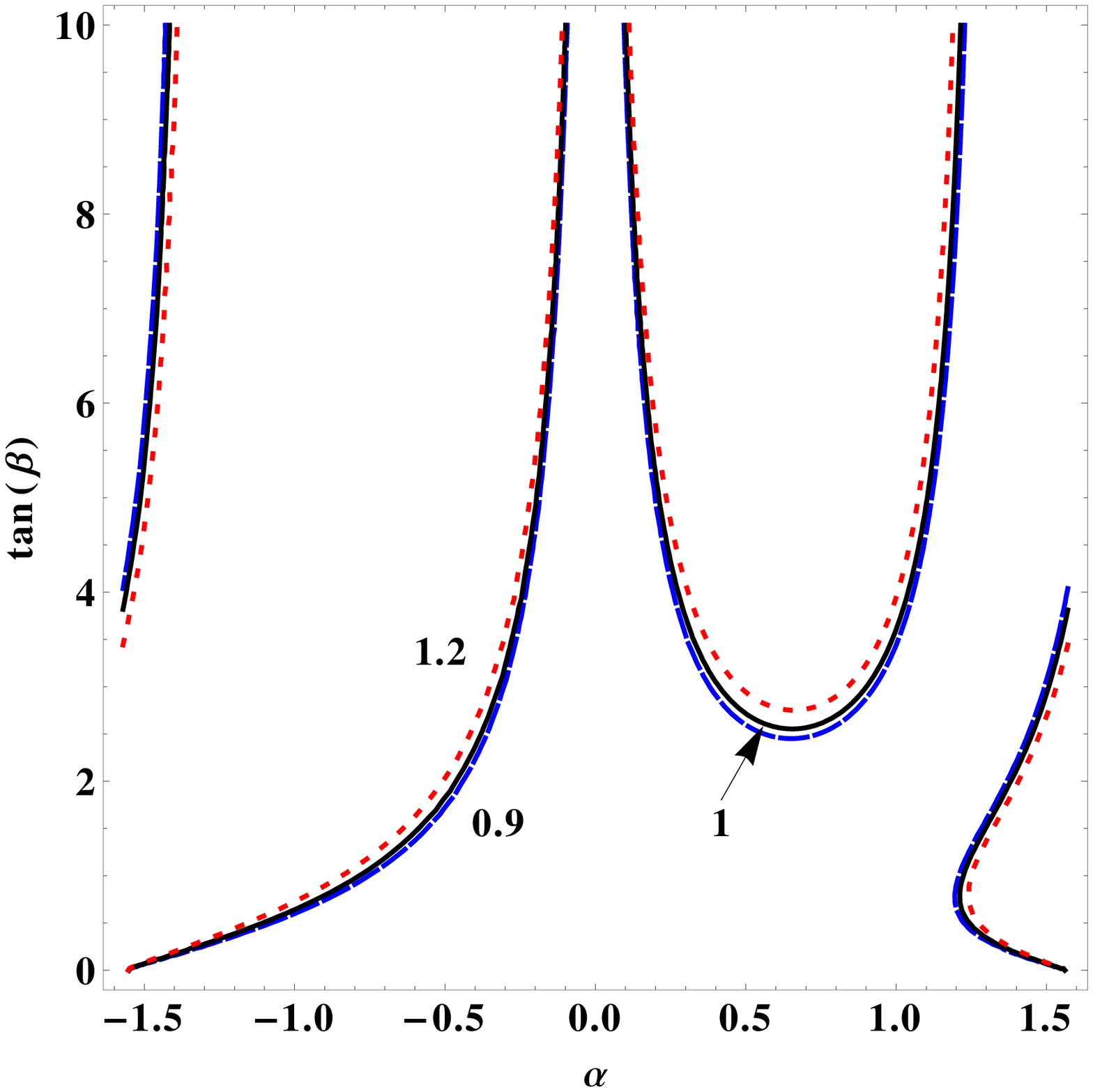}
}
\subfigure[]{
      \includegraphics[width=0.36\textwidth,angle=0,clip]{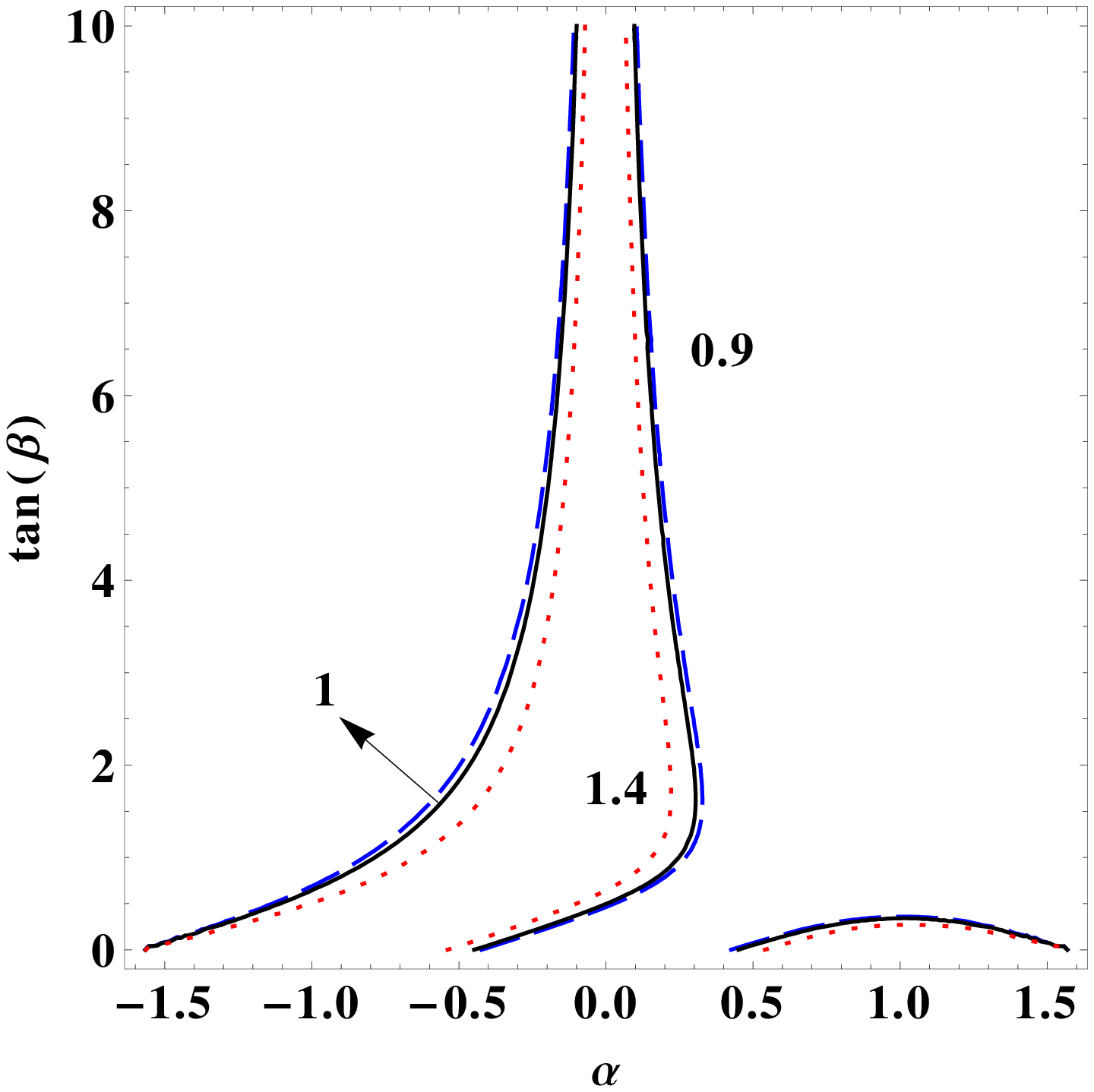}
}
\caption{
The contours correspond to constant  $R_{\tau\tau}^{VBF,Vh}$ as defined in the text.  For Type I (a): 
The blue (dashed), black (solid), and red (dotted)  lines are $R_{\tau\tau}^{VBF,Vh}=.9,1,1.1$, respectively.
 For Type II (b): 
The blue, black, and red lines are $R_{\tau\tau}^{VBF,Vh}=.9,1., 1.4$, respectively.
For the Lepton Specific model (c): 
The blue, black, and red lines are $R_{\tau\tau}^{VBF,Vh}=.9,1., 1.2$, respectively.
 For the Flipped Model(d): 
The blue, black, and red lines are $R_{\tau\tau}^{VBF,Vh}=.9,1., 1.4$, respectively.
}
\label{htt_fig}
\end{figure}

Preliminary LHC measurements of $h^0\rightarrow b {\overline b}$ 
and $h^0\rightarrow \tau^+\tau^-$ in the vector boson fusion (VBF) and  $V h^0$ channels are particularly interesting and
will serve to distinguish between the different 2HDMs.  Both VBF and 
$Vh^0$ production follow the same simple scaling between the Standard
Model  rates
and the 2HDM predictions, 
\begin{equation}
R_{ff}^{VBF,Vh} {\sim} \sin^2(\beta-\alpha)g_{hff}^2\, .
\end{equation} 
From Fig. \ref{hbb_fig}, it is clear that the Standard Model rates, $R_{bb}^{VBF,Vh}\sim 1$ only occurs for very specific values
of $\alpha$ and $\tan\beta$.  In Model I and the Lepton Specific model, $R_{bb}^{VBF,Vh}\sim 1.1$ 
requires $\tan\beta < 1$, while Model II and the Flipped model require $\alpha\sim -0.5$ and $\tan\beta > 4-8$ 
for $R_{bb}^{VBF,Vh}\sim 1.4$.  The rates to $\tau^+\tau^-$
normalized to the Standard Model predictions are shown in Fig. \ref{htt_fig}. 
Note that in Models I  and II,  $R_{bb}^{VBF,Vh} = R_{\tau\tau}^{VBF,Vh}$. This is because the $h^0$ coupling to 
$b{\overline b}$ is the same as the $h^0$ coupling to $\tau^+\tau^-$ in these two models,  
as shown in Table~\ref{table:coups}. As a result Fig.~\ref{hbb_fig}(a) is identical to Fig.~\ref{htt_fig}(a) and similarly 
Fig.~\ref{hbb_fig}(b) and Fig.~\ref{htt_fig}(b) are also identical to each other.
One can also see that Fig. 3(a) and Fig. 3(c) are very similar since the coupling $g_{hbb}$  
for Model I and the Lepton-Specific model are the same. They are not identical because the total widths in these two models are different.
This similarity also appears in Fig. 3(b) and Fig. 3(d) because the $h^0$ coupling to $b{\overline b}$ is the same
for Model II and the Flipped model.

\section{Review of Limits from Flavor Physics}
\label{bphys}
\begin{figure}[tb]
      \includegraphics[width=0.45\textwidth,angle=0,clip]{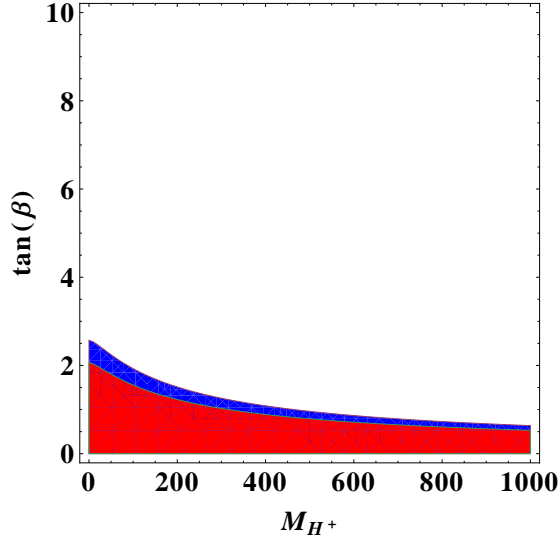}
\caption{Limits on 2HDMs from $\Delta M_{B_d}$.  Regions below the  blue and red regions are
excluded at $2$ and $3\sigma$, respectively.  
}
\label{dmb_fig}
\end{figure}

Limits on 2HDMs have been examined by many authors and we briefly update the results of Ref. \cite{Mahmoudi:2012ej}
using the SuperIso program\cite{Mahmoudi:2007vz, Mahmoudi:2008tp, Mahmoudi:2009zz} in order to examine the restrictions on $\tan\beta$ in the 2HDMs which are relevant
for Higgs coupling studies.  
Small values of $\tan\beta$ , $\tan\beta \lesssim 0.35 $, for $M_{H^+} \lesssim 2~TeV$ are excluded at $3\sigma$
 in all 2HDMs considered here by the experimental
measurement of $\Delta M_{B_d}$\cite{Barberio:2008fa},  
\begin{equation}
\Delta M_{B_d}\mid_{exp}=0.507\pm 0.004~ ps^{-1}\, .
\end{equation}
The limits from $\Delta M_{B_d}$ are identical in all 2HDMs studied in this work and the exclusion regions are shown in Fig. \ref{dmb_fig}.
For $M_{H^+}\sim 300 ~GeV$, $\tan\beta < 1.03$ is excluded at $3\sigma$. (For $M_{H^+}\sim 2~TeV$, $\tan\beta < .35$ is excluded at $3\sigma$.)

\begin{figure}[tb]
\subfigure[]{
      \includegraphics[width=0.45\textwidth,angle=0,clip]{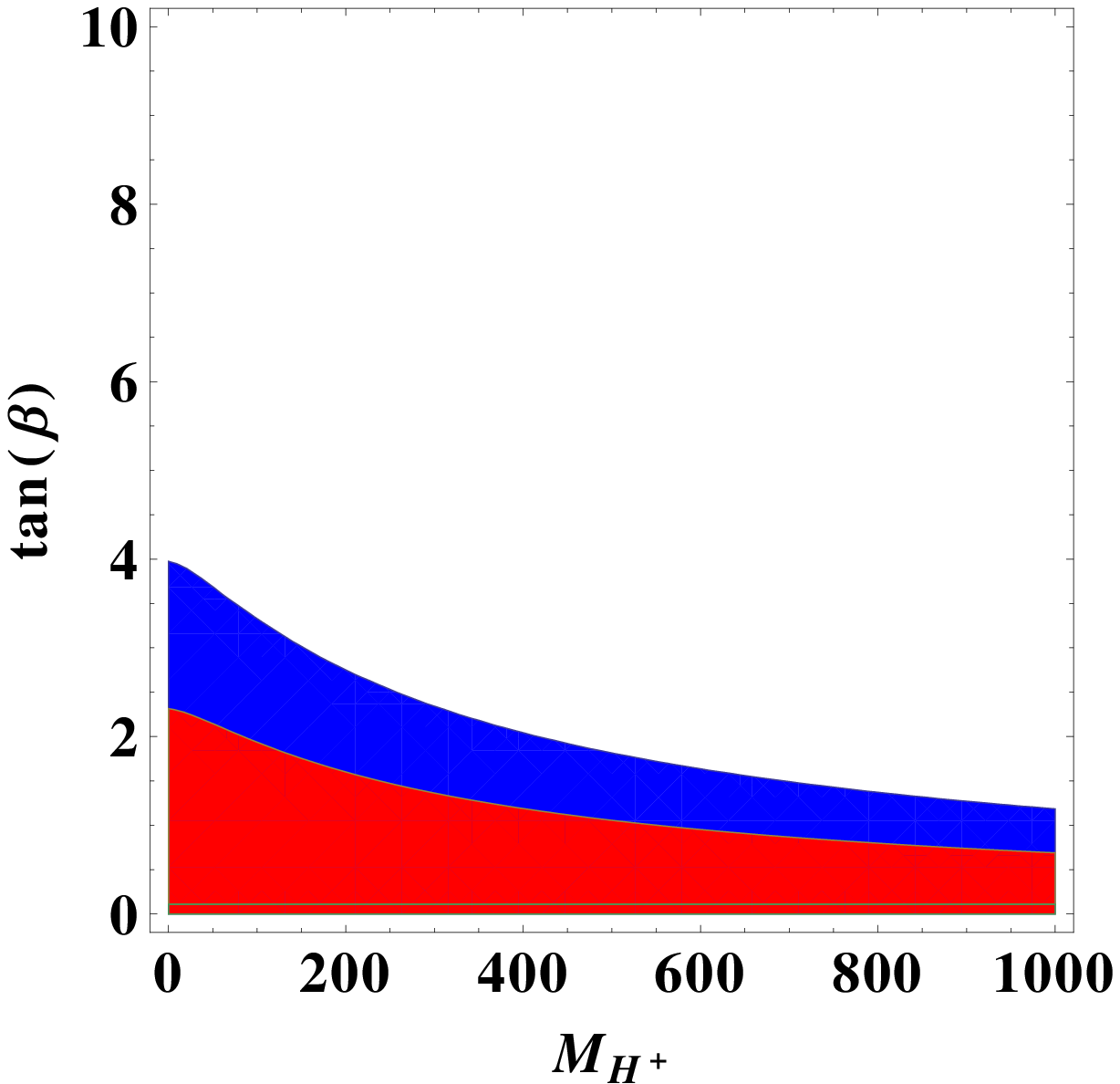}
}
\subfigure[]{
      \includegraphics[width=0.45\textwidth,angle=0,clip]{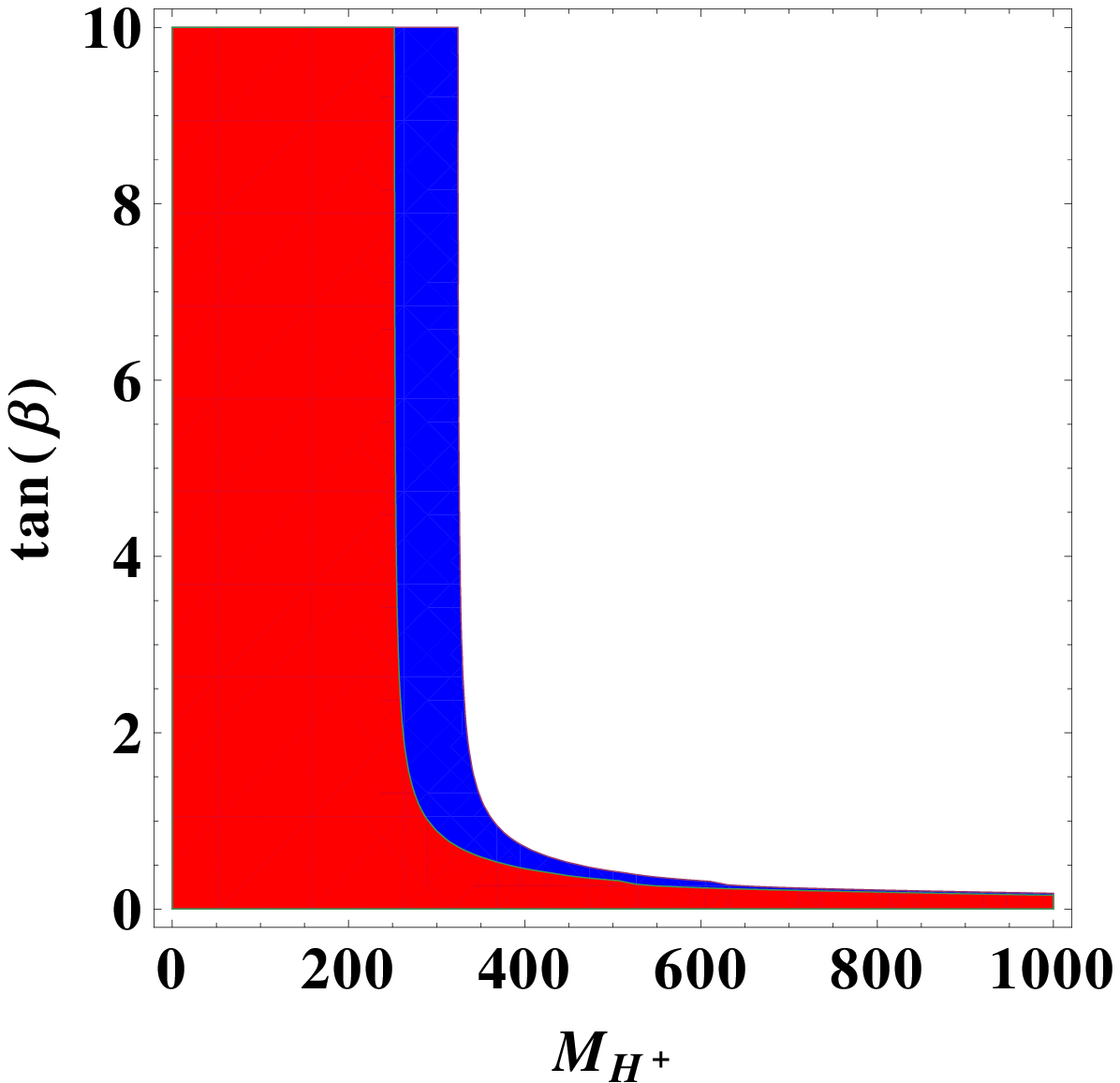}
}
\caption{Limits from the measurement of $B\rightarrow X_s \gamma$ in Type I and the Lepton Specific model (a). The regions
below the blue (upper) and red (lower) curves are excluded at $2$ and $3\sigma$. 
The limits in Type II and the Flipped Model are in (b) and the regions to the left of the blue and 
red curves are excluded at $2$ and $3\sigma$, respectively. }
\label{bsg_fig}
\end{figure}

Stringent bounds  also 
arise from the experimental measurement of
$B\rightarrow X_s\gamma$\cite{Barberio:2008fa,Aubert:2007my},
\begin{equation}
BR(B\rightarrow X_s \gamma)\mid_{exp} = (3.55\pm 0.24\pm 0.09)\times 10^{-4}\, ,
\end{equation}
and are shown in 
Fig. \ref{bsg_fig}\cite{Mahmoudi:2009zx,Mahmoudi:2012ej}.  For $M_{H^+}=1~TeV$, $\tan\beta < 0.69 (0.16)$ is excluded at $3\sigma$ in Type I and the Lepton
Specific model (Type II and the Flipped Model).  For $M_{H^+}=300~GeV$  the restrictions are still fairly stringent: $\tan\beta < 1.36 (0.88)$ is excluded 
at $3\sigma$ in Type I and the Lepton
Specific model (Type II and the Flipped Model).  For all values of $\tan\beta$, $M_{H^+}< 300~GeV$ is excluded in the Type II and Flipped models. 
The numerical results agree with Refs. \cite{Mahmoudi:2012ej,Baak:2011ze,Logan:2009uf,Logan:2010ag} in the appropriate limits.

Finally, the recent observation of the decay $B_s\rightarrow \mu^+\mu^-$\cite{:2012ct} 
has important implications for the 2HDMs,
\begin{equation}
BR(B_s\rightarrow\mu^+\mu^-)\mid_{exp}=(3.2^{+1.5}_{-1.2})\times10^{-9}\, ,
\end{equation}
in good agreement with the Standard Model prediction\cite{Buras:2012ru},
\begin{equation}
BR(B_s\rightarrow  \mu^+ \mu^-)\mid_{SM}=(3.23\pm 0.27)\times10^{-9}\, .
\end{equation}
New physics effects in $B_s\rightarrow \mu^+\mu^-$ come predominantly from the charged
Higgs exchanges.   
The contributions to $B_s\rightarrow \mu^+\mu^-$ in the type II two Higgs doublet model have been computed in the large $\tan\beta$ regime
in Refs. \cite{Logan:2000iv,Bobeth:2001sq,Ellis:2005sc},
 and adapted to the general Two Higgs doublet models considered here in the SuperIso program\cite{Mahmoudi:2008tp}.\footnote{
It is possible that there are contributions which are not enhanced by large $\tan\beta$ in the  Type II model  which
could be relevant in the Type I, Lepton Specific or Flipped models.  Such contributions are not
included in the SuperIso program.}

\begin{figure}[tb]
\subfigure[]{
     \includegraphics[width=0.5\textwidth,angle=0,clip]{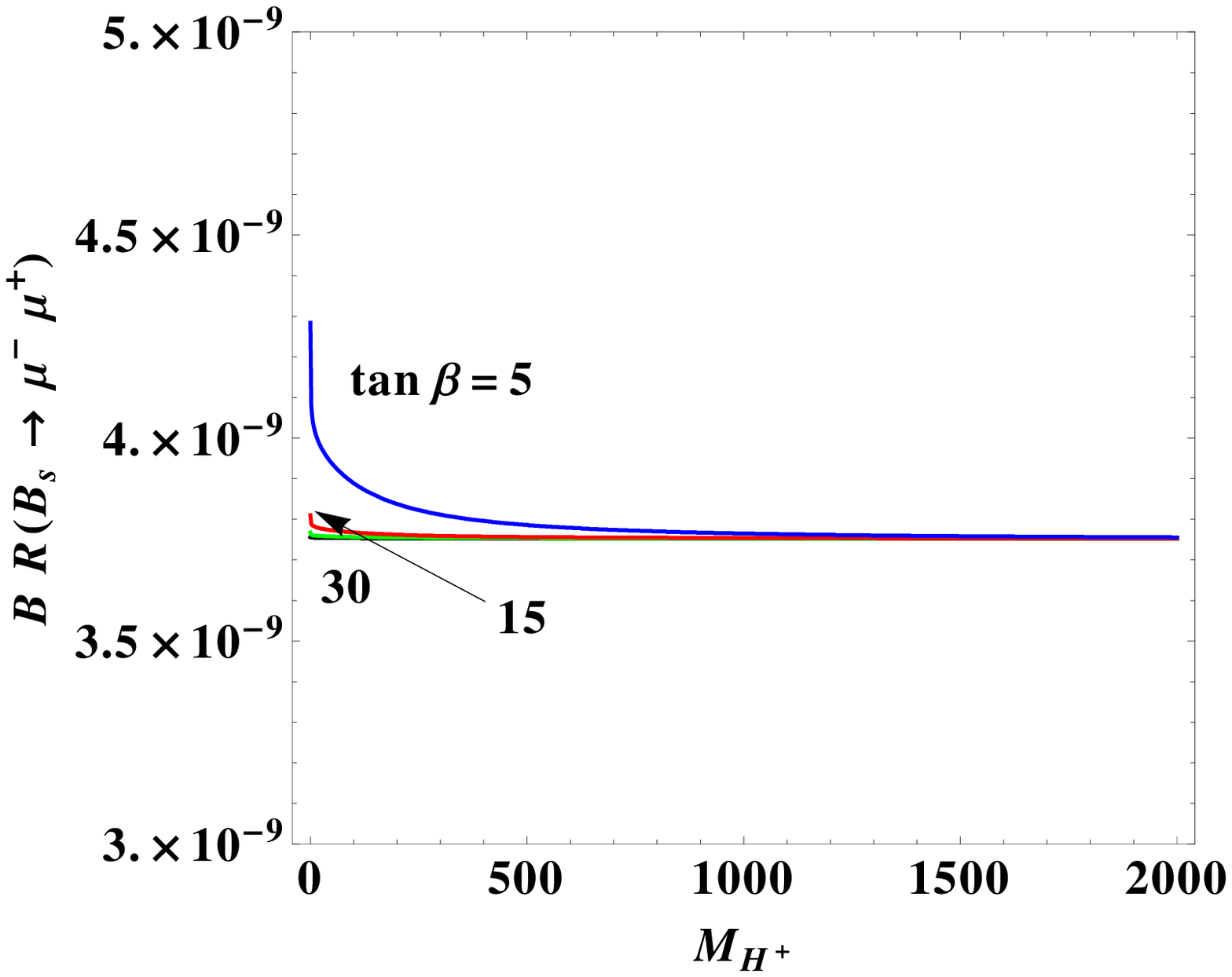}
}
\subfigure[]{
      \includegraphics[width=0.44\textwidth,angle=0,clip]{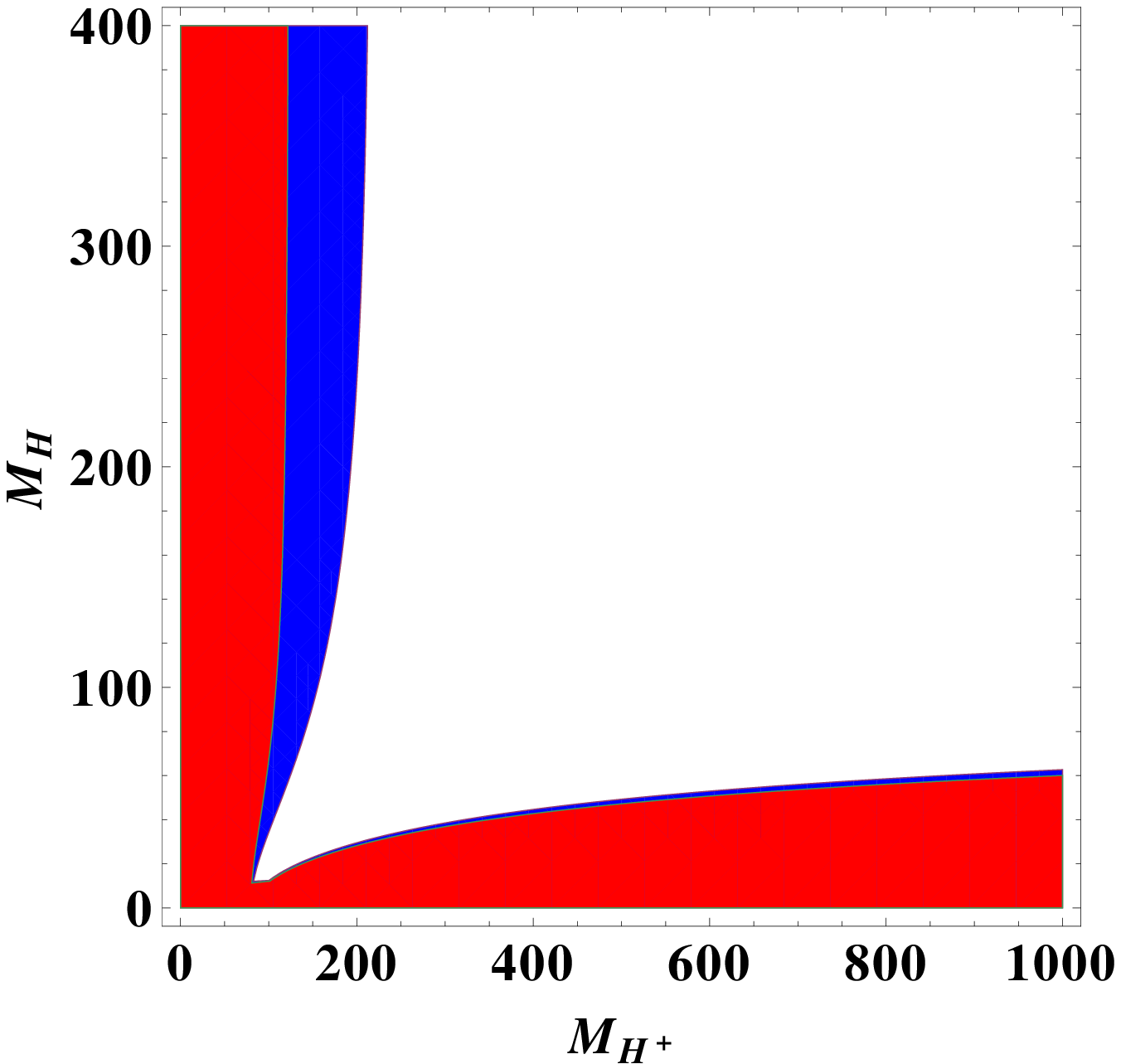}
}
\caption{Branching ratios for $B_s\rightarrow \mu^+\mu^-$ in Model I (a) with $M_{H^0}=M_A=300~GeV$
for $\tan\beta=$5 (blue), $15$(red),  and $30$ (green).  
The excluded region from $B_s\rightarrow \mu^+\mu^-$  in Model I is shown in  (b) with $M_{H^0}=M_{A}$. 
The red (blue+red) region of (b)  is excluded at $3(2) \sigma$ for $\alpha=-0.88$ and $\tan\beta=1$. These
values of $\alpha$ and $\tan\beta$ correspond to the best fit to the Higgs data which is derived in Section IV.  
This plot, however, is  
rather insensitive to the precise value of $\tan\beta$.}
\label{bmmt1_fig}
\end{figure}

The branching ratio for $B_s\rightarrow \mu^+\mu^-$ in the Type I model
is shown in Fig.\ref{bmmt1_fig}(a) as a function of $M_{H^+}$ for various values of $\tan\beta$.  
It is apparent that for $M_{H^+} > 500$ GeV, the branching ratio
is almost a constant, independent of $\tan\beta$.  In 
Fig.\ref{bmmt1_fig}(b), the excluded region is shown for the parameters which best fit the Higgs data (derived in
the next section).  However, for $M_{H^+}$ and $M_H > 500$ GeV,  the excluded region is not sensitive to the
value of $\tan\beta$.  Even for smaller values of $M_{H^+}$ and $M_H$, the sensitivity to
$\tan\beta$ is small.

\begin{figure}[tb]
\subfigure[]{
     \includegraphics[width=0.5\textwidth,angle=0,clip]{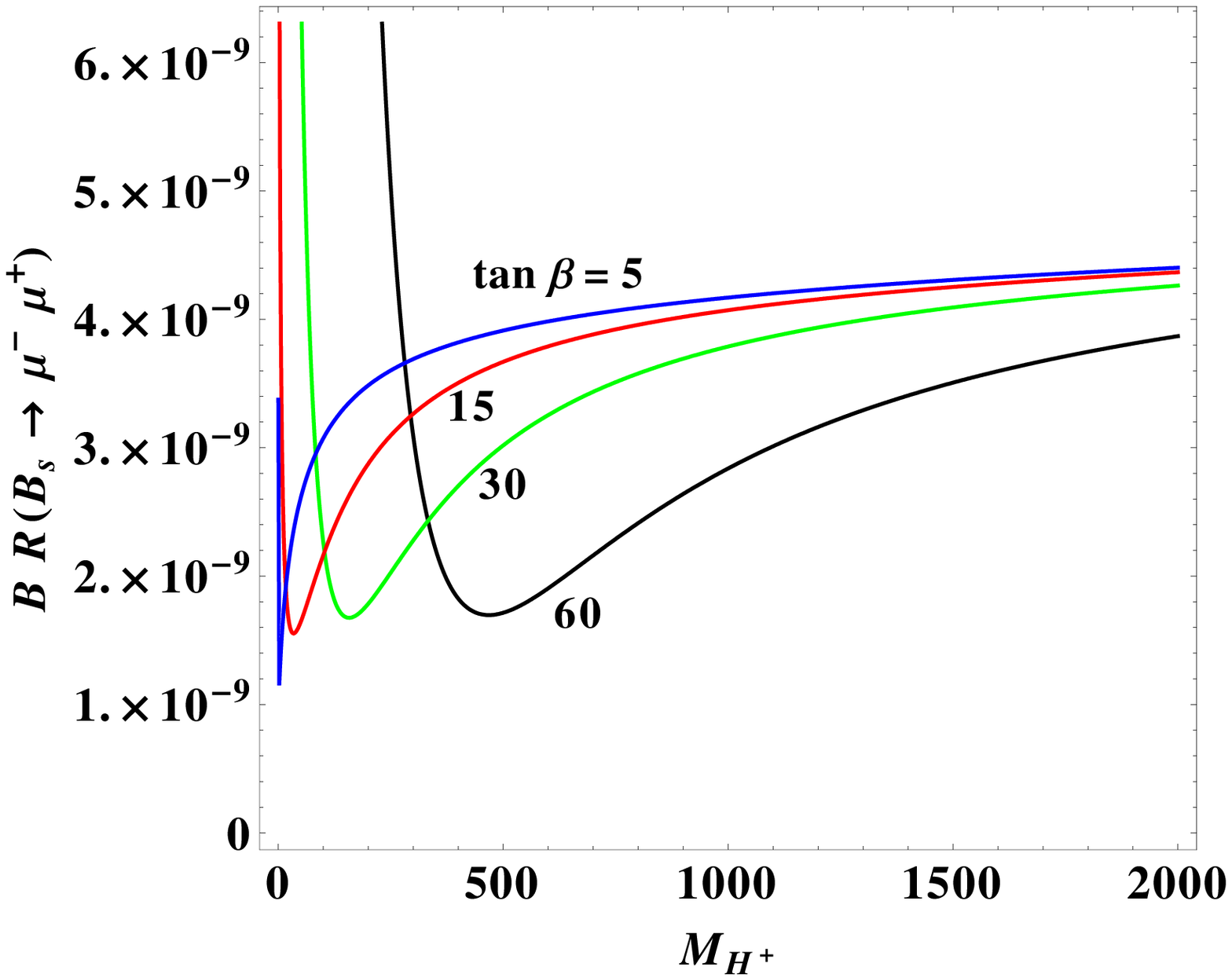}
}
\subfigure[]{
      \includegraphics[width=0.44\textwidth,angle=0,clip]{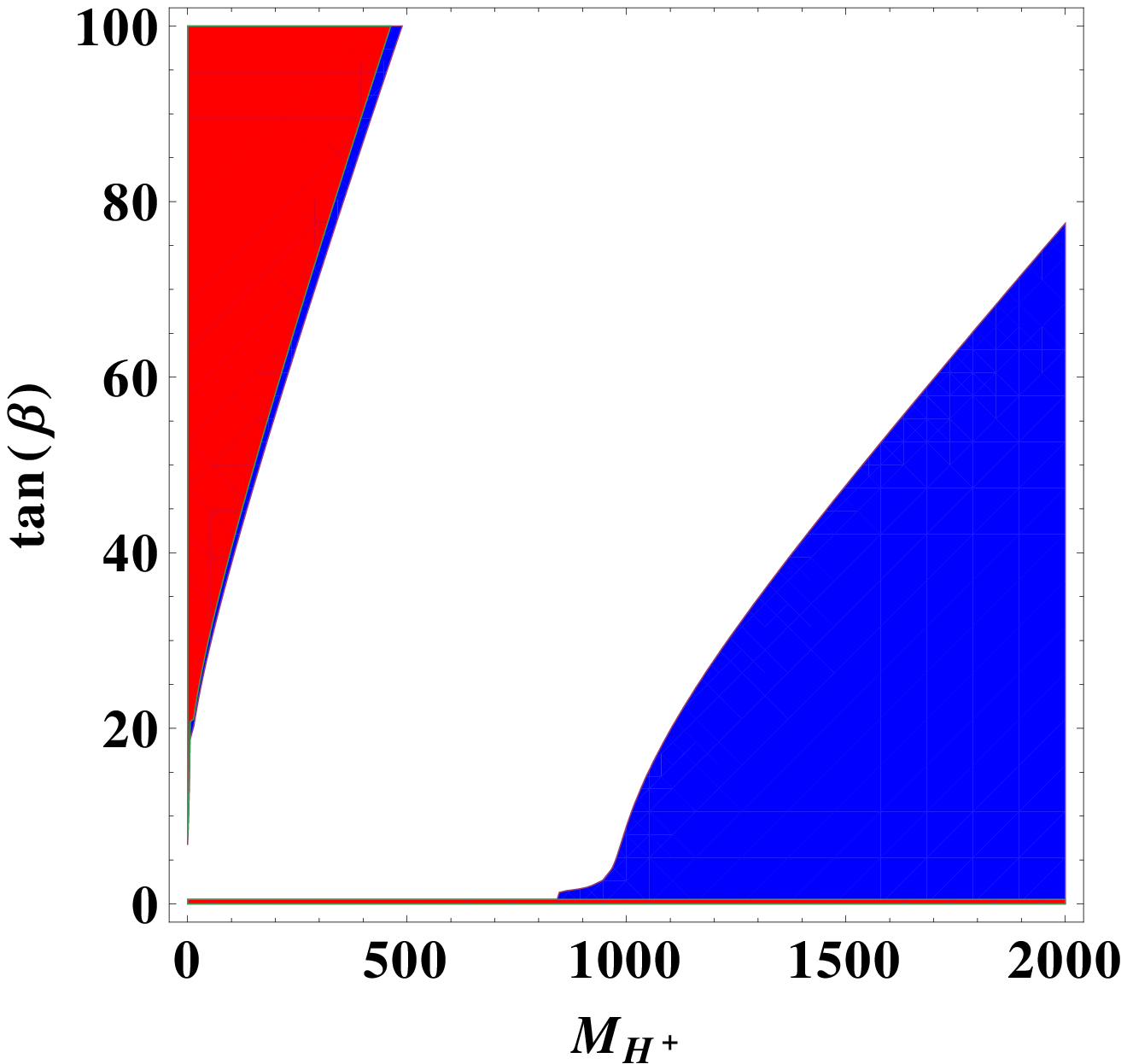}
}
\caption{Branching ratios for $B_s\rightarrow \mu^+\mu^-$ in Model II (a) with $M_{H^0}=M_A=300~GeV$
for $\tan\beta=$5 (blue), $15$(red), $30$ (green), and $60$ (black).  (At high $M_{H^+}$, $\tan \beta$ increases
going from the top to bottom curves.) 
The excluded region from $B_s\rightarrow \mu^+\mu^-$  in Model II  is shown in (b) with $M_{H^0}=M_{A}=145~GeV$. 
The red (blue+red) region of (b)  is excluded at $3(2) \sigma$.}
\label{bmm2_fig}
\end{figure}

\begin{figure}[tb]
\subfigure[]{
     \includegraphics[width=0.5\textwidth,angle=0,clip]{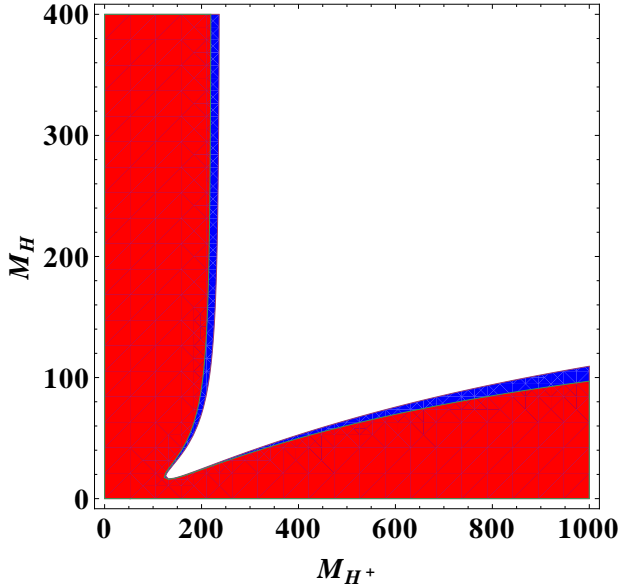}
}
\caption{
Excluded region from $B_s\rightarrow \mu^+\mu^-$  in Model II  with $M_{H^0}=M_{A}$
for $\alpha=-0.02$ and $\tan\beta=60$. (These values of $\alpha$ and $\beta$ correspond to the 
best fit to the Higgs data which is derived in Section IV.)
The red (blue+red) region of is excluded at $3(2) \sigma$.}
\label{bmmhh2_fig}
\end{figure}

  In the Type II model, the branching ratio $B_s\rightarrow \mu^+\mu^-$
has a significant dependence on $\tan\beta$ for small $M_{H^+}$, and goes to a constant
for very large $M_{H^+}$.  In this model 
  high values of $\tan\beta$ are excluded at $3\sigma$ for
small $M_{H^+}$, while there is a $2\sigma$ excluded region at large $\tan\beta$, as shown in Fig.
\ref{bmm2_fig}.  The dependence of the excluded region  on the choice of neutral Higgs masses, $M_{H^0}$ and $M_A$,  is
shown in Fig. \ref{bmmhh2_fig} for $\alpha=-0.02$ and $\tan\beta=60$, (which corresponds to the best fit to the Higgs
data derived in Section IV).

\begin{figure}[tb]
\subfigure[]{
      \includegraphics[width=0.5\textwidth,angle=0,clip]{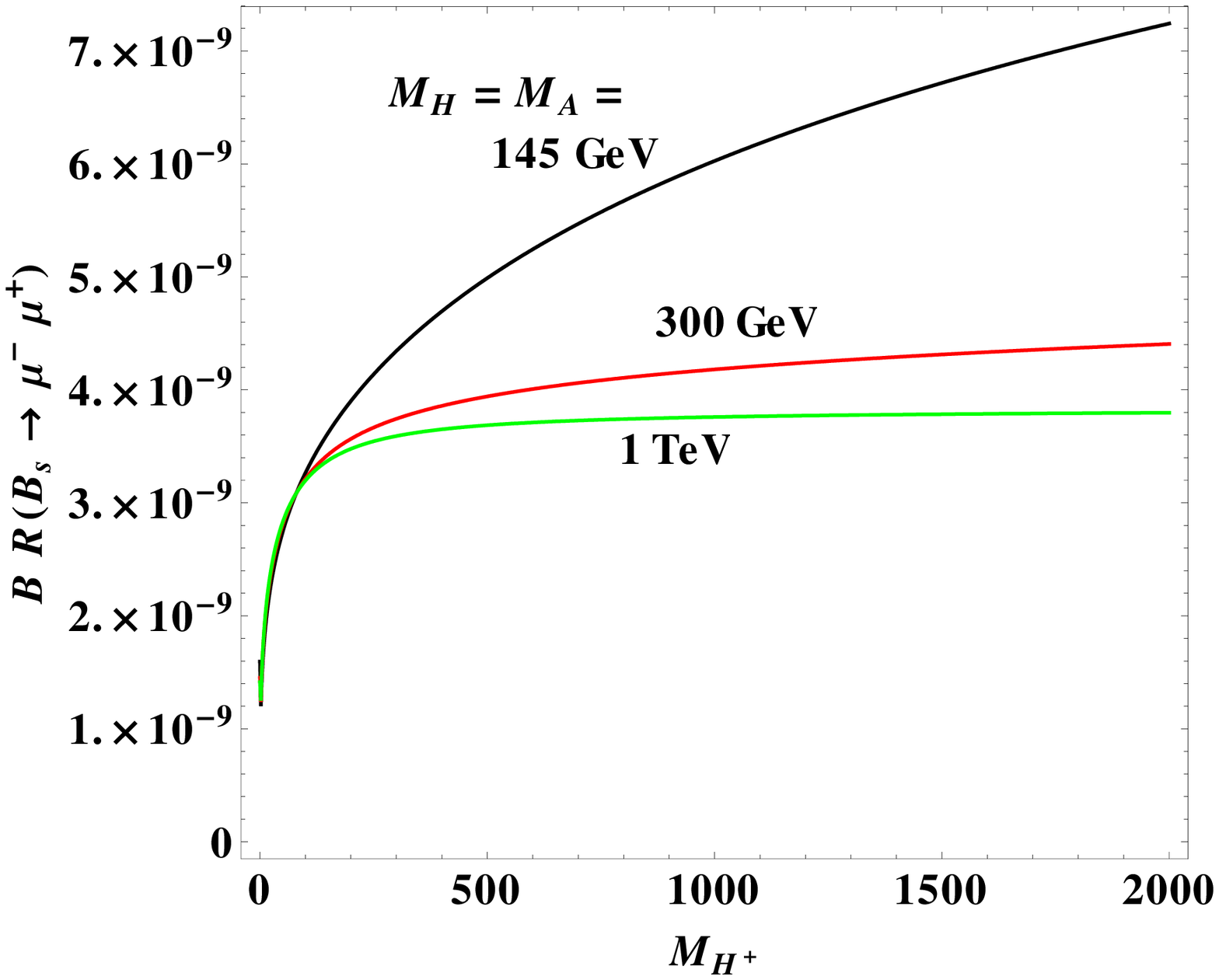}
}
\subfigure[]{
    \includegraphics[width=0.45\textwidth,angle=0,clip]{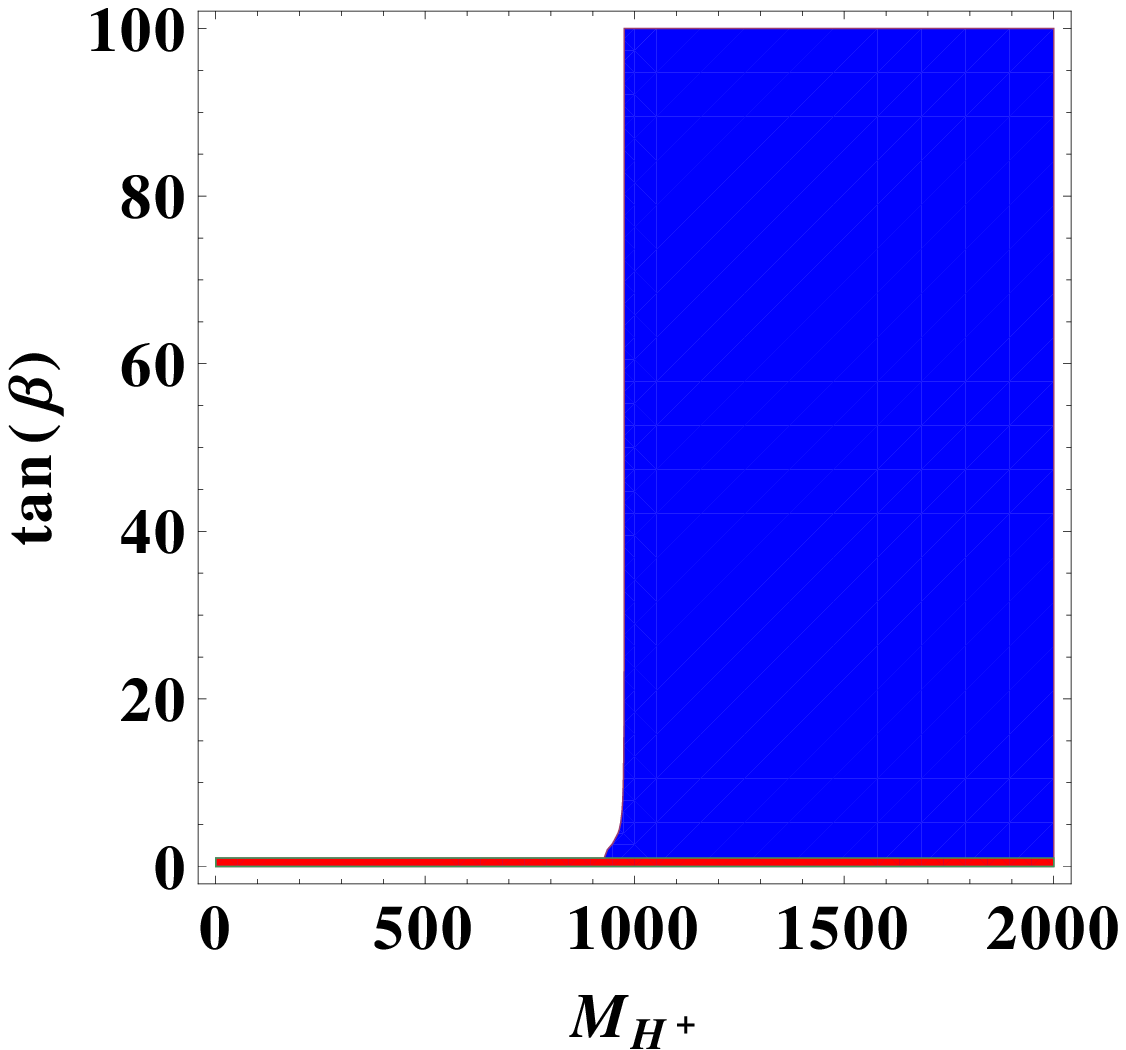}
}
\caption{
Branching ratios for $B_s\rightarrow \mu^+\mu^-$  in the Lepton Specific model (a).  (The
branching ratio is almost independent of
 $\tan\beta$.)   The excluded region for $M_{H^0}=M_A=145~GeV$ is shown in (b). 
The red (blue+red) region of is excluded at $3(2) \sigma$.
}
\label{bmmls_fig}
\end{figure}

\begin{figure}[tb]
\subfigure[]{
     \includegraphics[width=0.5\textwidth,angle=0,clip]{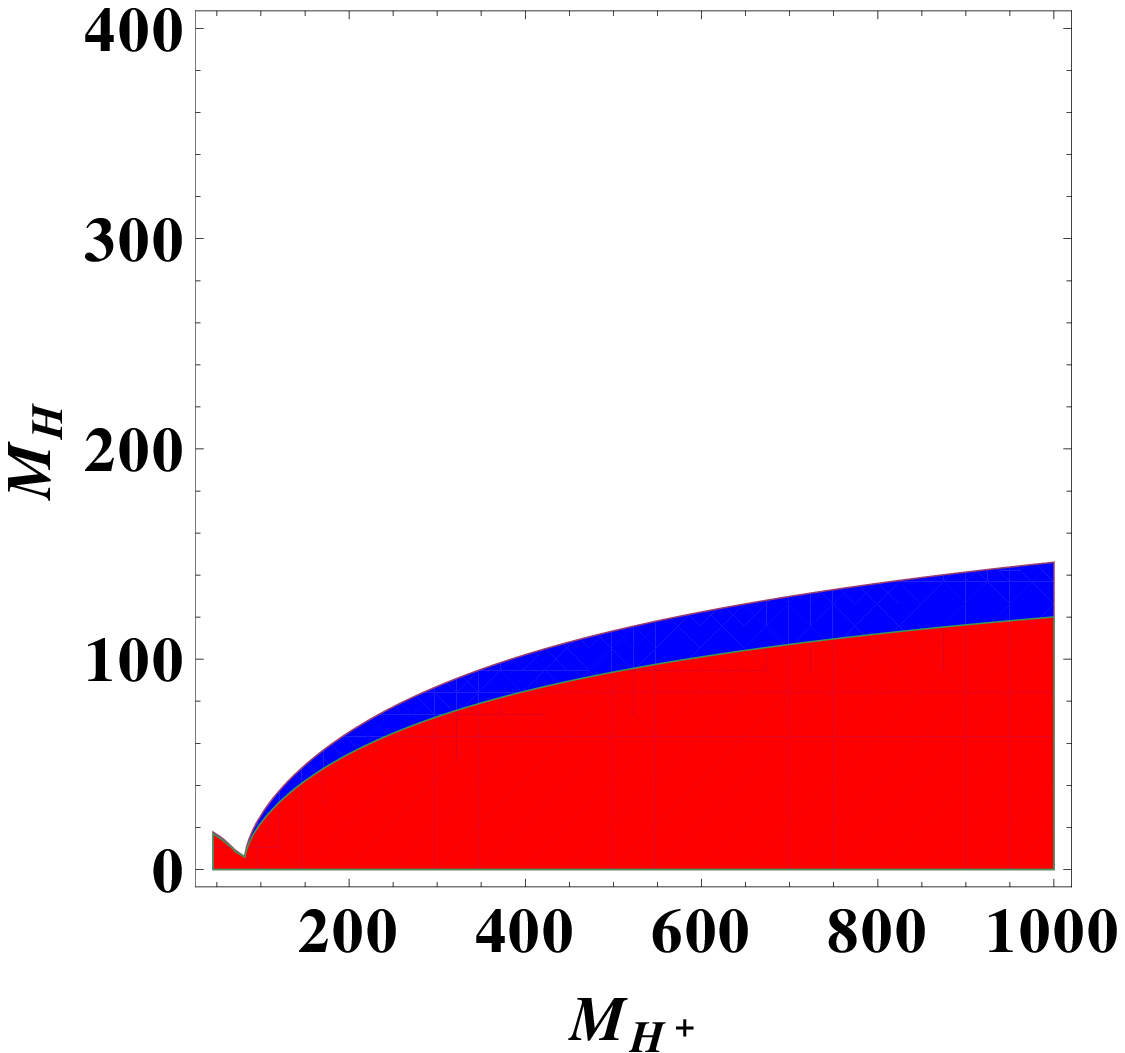}
}
\caption{
Excluded region from $B_s\rightarrow \mu^+\mu^-$  in the Lepton Specific model  with $M_{H^0}=M_{A}$ and
$\alpha=-0.02$, the best fit to the Higgs data, derived in Section IV.  
(The excluded region is independent of $\tan\beta$.) 
The red (blue) region of is excluded at $3(2) \sigma$.}
\label{bmmhhls_fig}
\end{figure}

 For the Lepton Specific case, the  dominant contribution 
to $B_s\rightarrow \mu^+\mu^-$ is proportional
to $\lambda_{tt}\lambda_{\mu\mu}$  and so the branching ratio is insensitive to $\tan\beta$.
In Fig. \ref{bmmls_fig}, we show the $BR(B_s\rightarrow \mu^+\mu^-)$ as a function of $M_{H^+}$ and 
the regions which are excluded at $2$ and $3\sigma$ from this decay.  We see that the excluded
region does not depend on $\tan\beta$.  For heavy charged Higgs masses,
the branching ratio approaches a constant.  The dependence on the choice of neutral Higgs masses is shown
in Fig. \ref{bmmhhls_fig}. 
From Fig. \ref{bmmls_fig}(a), it is clear that the branching ratio for $B_s\rightarrow \mu^+\mu^-$ 
increases for $M_{H^+} \gtrsim 150$ GeV as $M_{H^0}$ and $M_A$ decrease. 
This leads to the exclusion of the region between $M_{H^0}=M_A \lesssim 100$ GeV and $M_{H^+} \gtrsim 150$ GeV at 2 
$\sigma$ in Fig. \ref{bmmhhls_fig}. 

\begin{figure}[tb]
\subfigure[]{
     \includegraphics[width=0.5\textwidth,angle=0,clip]{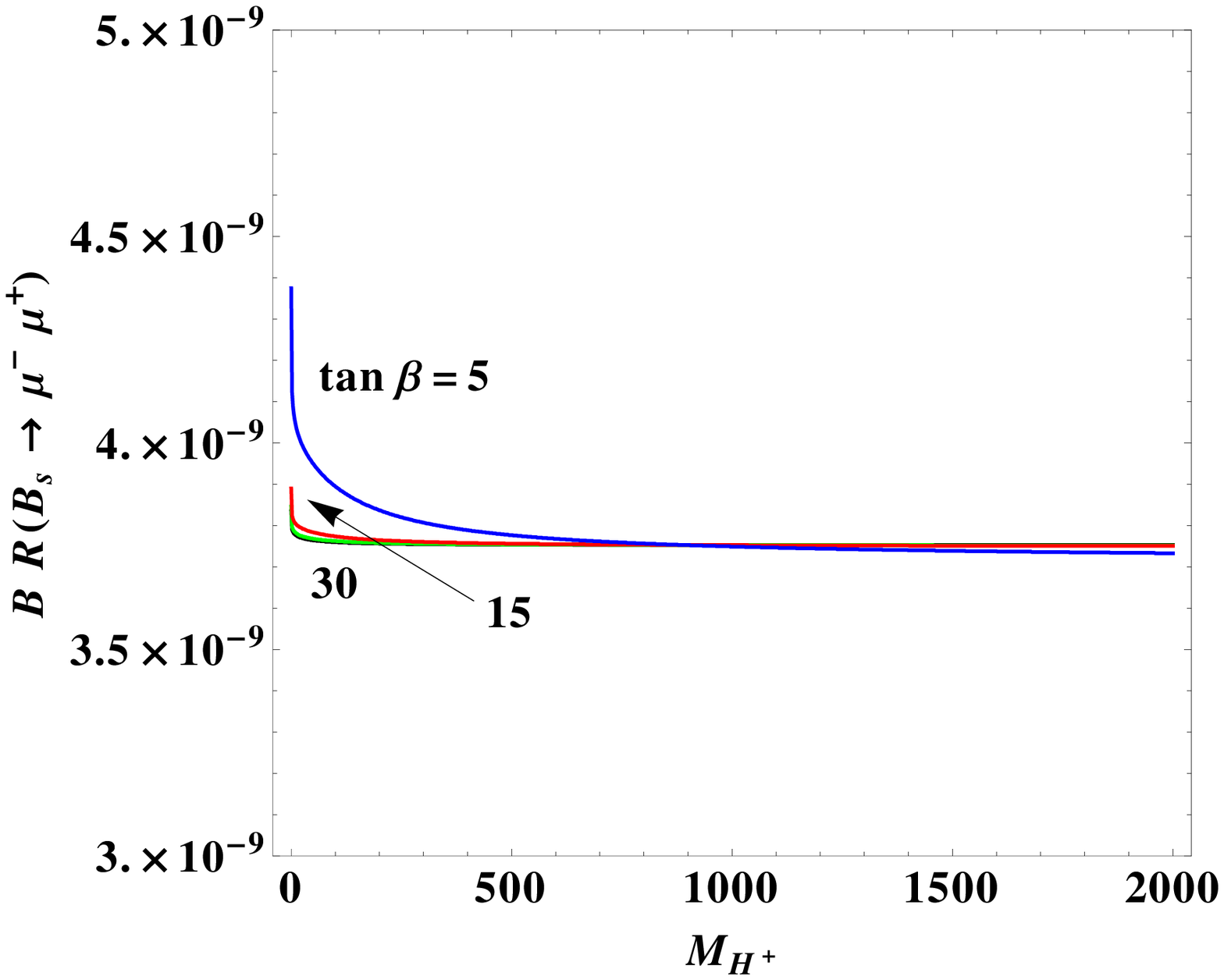}
}
\subfigure[]{
      \includegraphics[width=0.44\textwidth,angle=0,clip]{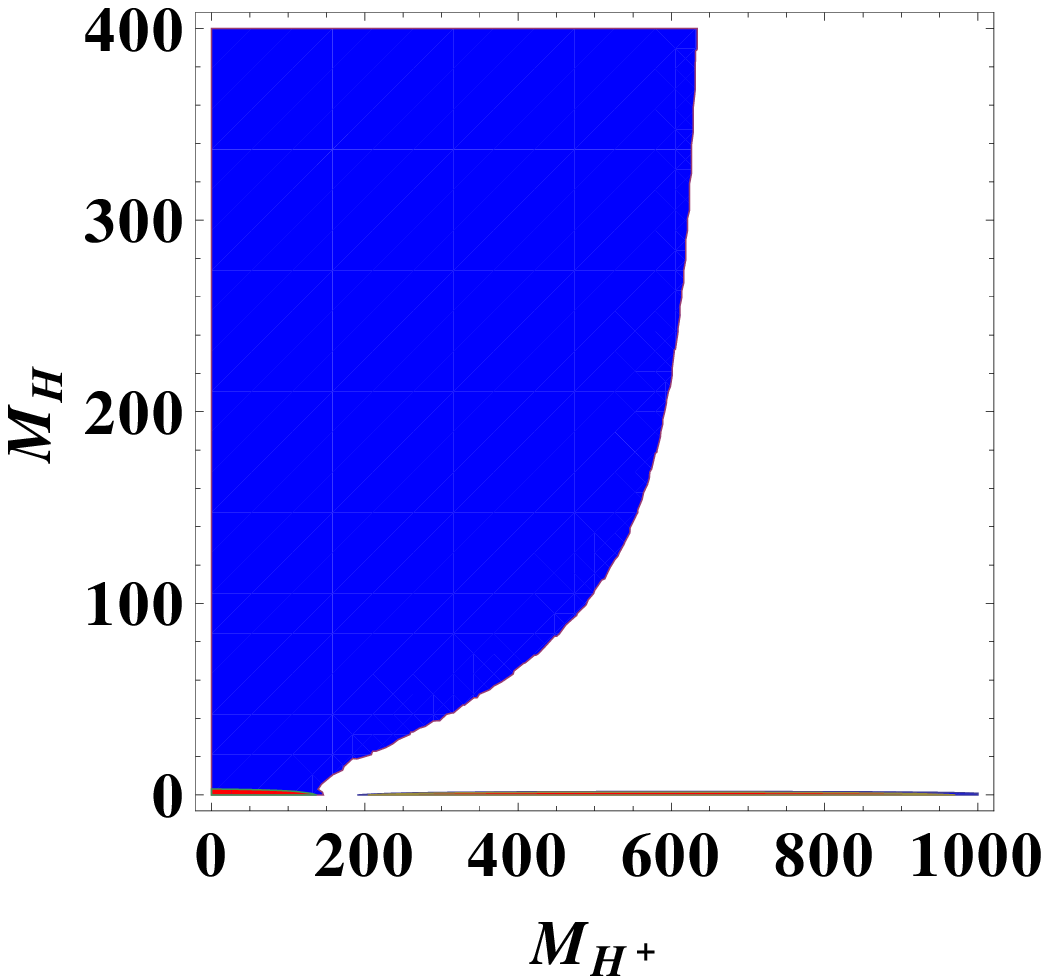}
}
\caption{Branching ratios for $B_s\rightarrow \mu^+\mu^-$ in the Flipped model (a) with $M_{H^0}=M_A=300~GeV$
for $\tan\beta
=$5 (blue), $15$(red),  and $30$ (green).  
Excluded region from $B_s\rightarrow \mu^+\mu^-$  in the Flipped model (b) (with $M_{H^0}=M_{A}$). 
The blue region of (b)  is excluded at $2 \sigma$ if we assume the uncertainty is five times smaller than 
the current one for $\alpha=-0.01$ and $\tan\beta=76.56$, which corresponds to the best fit to the Higgs data
derived in Section IV.}
\label{bmmfl_fig}
\end{figure}

The branching ratio for $B_s\rightarrow \mu^+\mu^-$ for the Flipped model is shown in Fig.~\ref{bmmfl_fig} for several
values of  $\tan\beta$ and for $M_{H^0}=M_A=300~GeV$.  For $M_{H^+}>300~GeV$, the branching ratio is insensitive
to the input parameters.  
 
\section{Results from Higgs Measurements}
\begin{table}[tp]
\renewcommand{\arraystretch}{1.4}
\caption{Measured Higgs Signal Strengths}
\centering
\begin{tabular}{|c|c|c|}
\hline
 Decay          & Production & Measured Signal Strength $R^{meas}$ \\ \hline
%\multirow{7}{*}
{$\gamma \gamma$} 
                & ggF        & $1.8 \pm 0.4\pm 0.2\pm 0.2$, [ATLAS] \cite{atlas-168}\\ 
                & VBF        & $2.0 \pm 1.2 \pm0.6 \pm 0.3$  [ATLAS]\cite{atlas-168}\\ 
                & inclusive  & $1.8 \pm 0.4$   [ATLAS]\cite{atlas-170}\\ 
                & ggF        & $1.4 \pm 0.6$   [CMS]\cite{cms-020}\\ 
                & VBF        & $2.1^{+1.4}_{-1.1}$  [CMS]\cite{cms-020}\\ 
                & inclusive  & $1.56 \pm 0.43$      [CMS]\cite{cms-020}\\
                & ggF        & $6.1^{+3.3}_{-3.2}$  [Tevatron]\cite{hcp:Enari}\\ \hline
%                
%\multirow{5}{*}
{$W W$}             
                & ggF        & $1.5 \pm 0.6$          [ATLAS] \cite{atlas-170}\\ 
                & ggF        & $0.74 \pm 0.25$        [CMS]\cite{cms-042}\\ 
                & VBF        & $0.3^{+1.5}_{-1.6}$    [CMS]\cite{cms-020}\\ 
                & Wh         & $-2.9^{+3.2}_{-2.9}$   [CMS]\cite{cms-020}\\ 
                & ggF        & $0.8^{+0.9}_{-0.8}$    [Tevatron]\cite{hcp:Enari}\\ \hline
%
%\multirow{2}{*}
{$ZZ$}              
                & inclusive        & $1.0 \pm 0.4$         [ATLAS]\cite{atlas-170}\\ 
                & inclusive        & $0.8^{+0.35}_{-0.28}$ [CMS]\cite{cms-041}\\ \hline
\end{tabular}

\vspace{-1ex}
%\caption{$*$ Only 8 TeV results are used in this channel.
%}
\label{tab:models1}
\end{table}

\begin{table}[tp]
\renewcommand{\arraystretch}{1.4}
\caption{Measured Higgs Signal Strengths}
\centering
\begin{tabular}{|c|c|c|}
\hline
 Decay          & Production & Measured Signal Strength $R^{meas}$ \\ \hline
%
%\multirow{5}{*}
{$b\bar{b}$}      
                & Vh         & $-0.4 \pm 1.0$    [ATLAS] \cite{atlas-170}\\ 
                & Vh         & $1.3^{+0.7}_{-0.6}$      [CMS]\cite{cms-044}\\ 
                & Vh         & $1.56^{+0.72}_{-0.73}$   [Tevatron]\cite{hcp:Enari}\\ \hline
%\multirow{8}{*}
{$\tau^+ \tau^-$}
                & ggF        & $2.4 \pm 1.5$          [ATLAS]\cite{atlas-160}\\ 
                & VBF        & $-0.4 \pm 1.5$         [ATLAS]\cite{atlas-160}\\ 
                & inclusive  & $0.8 \pm 0.7$          [ATLAS]\cite{atlas-170}\\ 
                & ggF        & $0.9^{+0.8}_{-0.9}$    [CMS]\cite{cms-043}\\ 
                & VBF        & $0.7 \pm 0.8$          [CMS]\cite{cms-043}\\ 
                & Vh         & $1.0^{+1.7}_{-2.0}$    [CMS]\cite{cms-043}\\ 
                & inclusive  & $0.72 \pm 0.52$        [CMS]\cite{cms-043}\\ 
                & ggF        & $2.1^{+2.2}_{-1.9}$    [Tevatron]\cite{hcp:Enari}\\ \hline                 
\end{tabular}

\vspace{-1ex}
\label{tab:models2}
\end{table}

\begin{figure}[tb]
\subfigure[]{
      \includegraphics[width=0.36\textwidth,angle=0,clip]{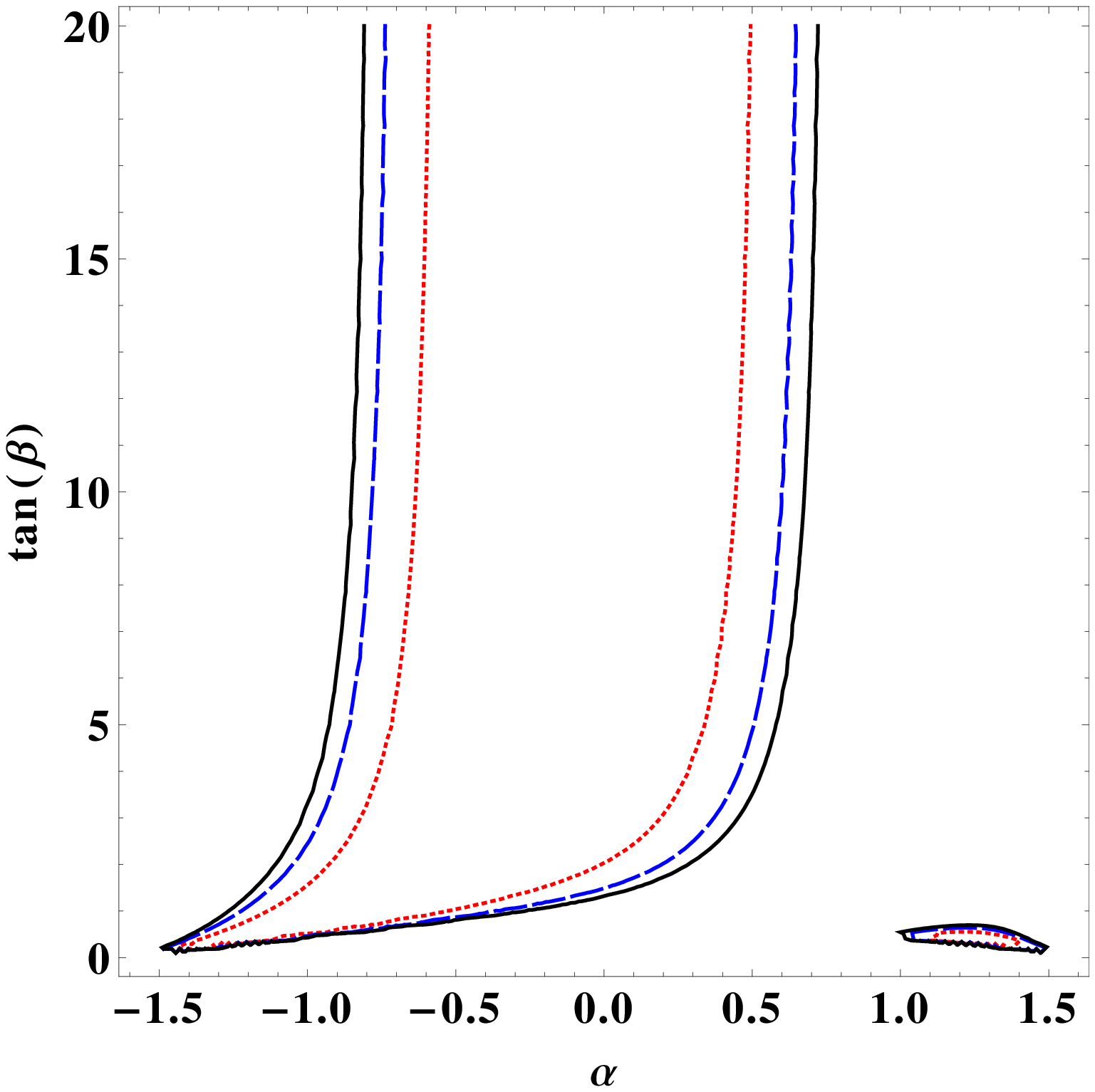}
}
\subfigure[]{
      \includegraphics[width=0.36\textwidth,angle=0,clip]{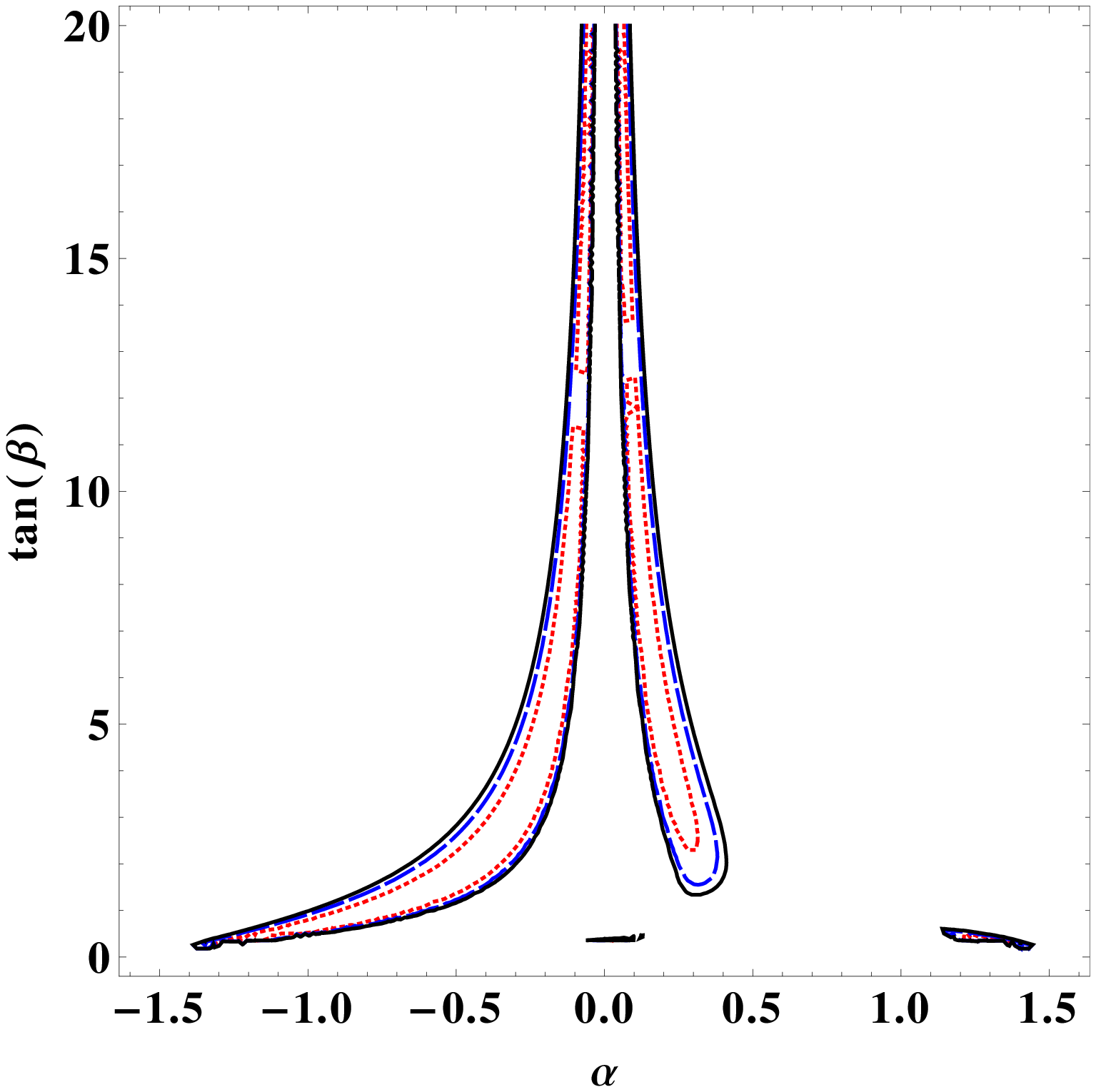}
}
\subfigure[]{
      \includegraphics[width=0.36\textwidth,angle=0,clip]{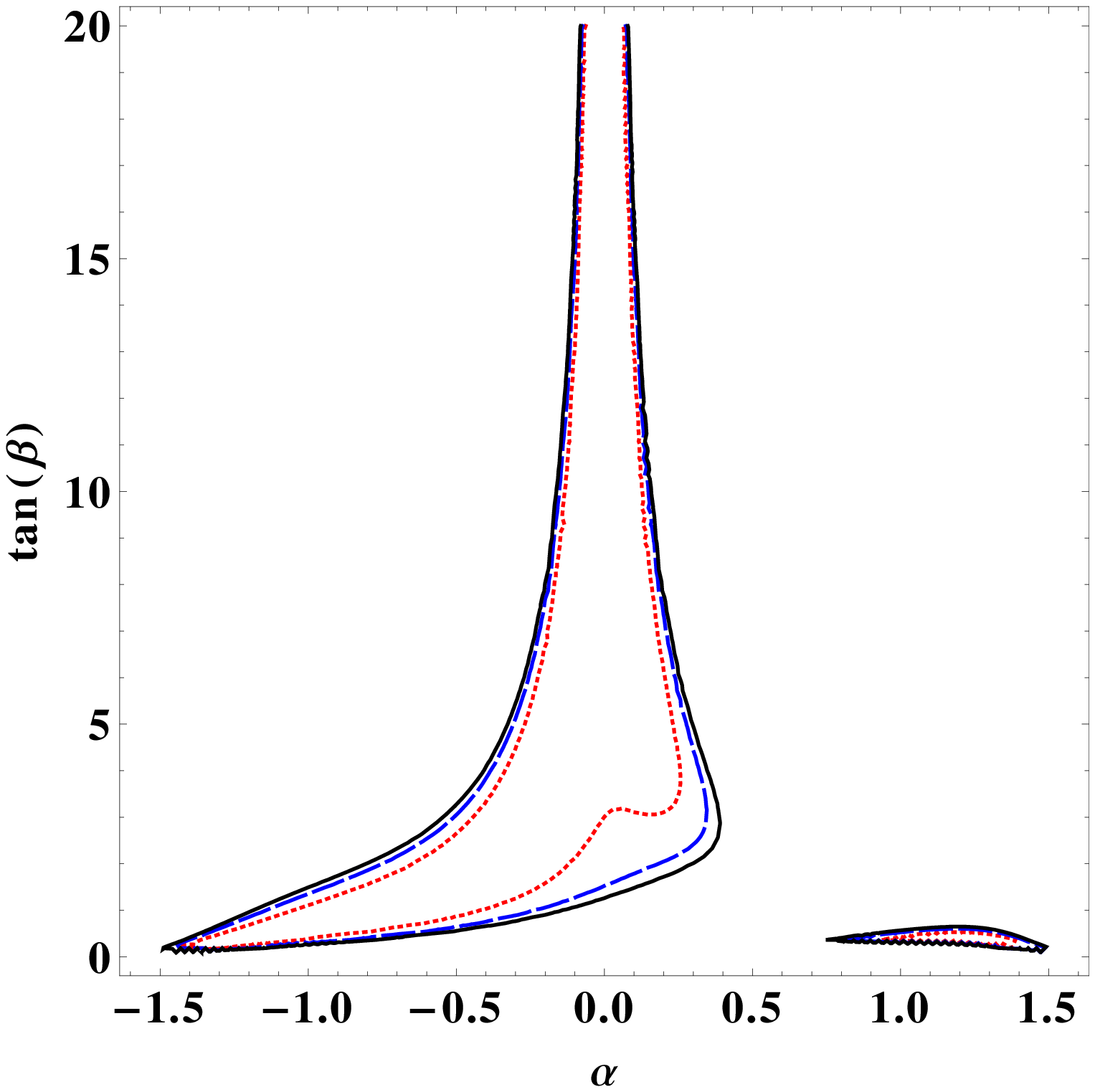}
}
\subfigure[]{
      \includegraphics[width=0.36\textwidth,angle=0,clip]{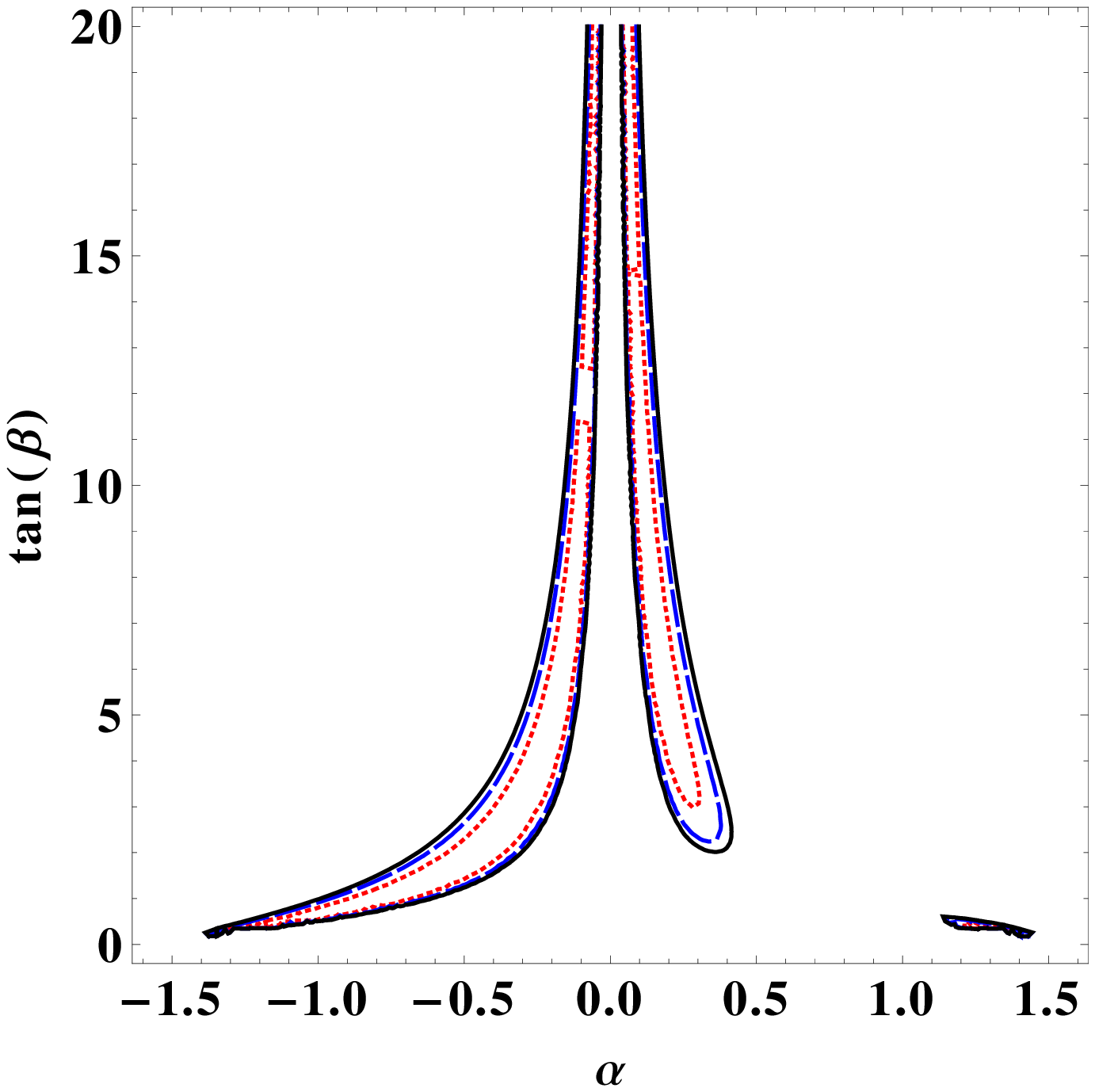}
}
\caption{Allowed regions  in the $\alpha-\tan\beta$ plane  in Type I (a), Type II (b), Lepton Specific (c),
and Flipped (d) 2HDMs obtained by minimizing the $\chi^2$ with no restrictions from flavor physics.  
The region between the black (solid), blue (dashed), and
red (dotted) lines is allowed at $99\%, ~95\%$, and $68\%$ confidence level.
}
\label{chisq_fig}
\end{figure}

We do a simple $\chi^2$ fit to the data shown in Tables \ref{tab:models1} and \ref{tab:models2}  assuming $M_{h^0}=125~GeV$.
We follow the standard definition of $\chi^2 = \Sigma_i {(R_i^{\rm 2HDM}-R_i^{\rm meas})^2\over (\sigma^{meas}_i)^2}$, where $R^{2HDM}$ 
represents predictions for the signal strength from the 2HDMs and $R^{\rm meas}$ stands for the most recent results of the measured 
signal strength shown in Tables \ref{tab:models1} and \ref{tab:models2} by the ATLAS and CMS collaborations at the LHC. 
$\sigma^{meas}$ denotes the uncertainty of $R^{meas}$.\footnote{The VBF tagged channels have a small contribution from gluon fusion.
The ATLAS results (Table 1 of Ref. \cite{atlas-168}) explicitly separate the true VBF contribution from the gluon fusion channel using
Monte Carlo.  The CMS VBF results of
Ref. \cite{cms-045} (Table 1)  contain a $30-50\%$ contamination from gluon fusion, estimated from Monte Carlo.  We
assume a $30\%$ contamination of the CMS VBF result from gluon fusion, although the
 results of Fig. \ref{chisq_fig} are not sensitive to this assumption.}
The results  (with no constraints from flavor physics) are shown in Fig. \ref{chisq_fig}. In all cases, the $\chi^2$ minima occurs
for $\alpha\sim 1.3-1.4$ and $\tan\beta \sim 0.2-0.3$. The results of Sec. \ref{bphys}, however,
 show that even for heavy $M_{H^+}$ and heavy
$M_{H^0}$ and $M_A$, such small values of $\tan\beta$ are not allowed in the 2HDMs we consider here due to constraints
from the $B$ sector.  

We perform a constrained fit to the data requiring $\tan\beta > 1$, which is consistent with $B$ physics
data of the previous section.  In all models, the results  of Fig. \ref{chisq_fig}
show that there are large regions of parameter space allowed at both the $2$ and $3\sigma$ confidence levels
and the results have only a mild dependence on $\tan\beta$.  In Model II and the Flipped model, $\alpha=0$ is not
allowed, primarily due to the $h^0\rightarrow b {\overline b}$ measurement.   In Model II, the Lepton Specific, and
the Flipped model, only a small range of $\alpha$ is allowed and the $\chi^2$ minimum occurs for
large $\tan\beta$ ($60, 54$, and $77$, respectively). In Model I, a fairly large range of $\alpha$ is
consistent with the data and the $\chi^2$ minimum is at $\tan\beta=1$.

\section{Conclusions}

We have considered four variations of 2HDMs which have a $Z_2$ symmetry suppressing tree level FCNCs.  
Higgs production and decay in the 2HDMs can be significantly different than in the Standard Model and only
small regions of $\alpha-\tan\beta$ can produce rates which are consistent with the experimental results from the
LHC.  Further, the parameters
of these models are strongly constrained by measurements in the $B$ sector.
In particular, limits on $\Delta M_{B_d}$  require $\tan\beta \gtrsim 0.35$ for $M_{H+} \lesssim 2 ~TeV$ in all
2HDMs considered here.  For each model, we have also shown the regions in parameter space which are allowed by
the measurement of $B_s\rightarrow \mu^+\mu^-$ for the parameters which correspond to the best fit to the
Higgs data.
Unitarity also restricts the allowed regions to have $\tan\beta \gtrsim 0.28$.
Our major result is shown in Fig. \ref{chisq_fig} where we show the regions of $\alpha-\tan\beta$ which are consistent
with the Higgs cross section and branching ratio
 results at the $2$ and $3\sigma$ level.  None of the models we studied can be excluded by current measurements. 
\section*{Acknowledgements}
We would like to thank Alejandro Celis, Nathaniel Craig, Kyle Cranmer, Jamison Galloway, Marc-Andre Pleier, Gabe Shaughnessy, 
and Scott Thomas for useful discussions.
We also thank F. Mahmoudi for help with SuperIso. This work is supported by the United States Department of Energy under
Grant DE-AC02-98CH10886.
\bibliography{twohiggs}
\end{document}